\documentclass[prd,showpcs,amsmath,amssymb,nofootinbib,longbibliography,twocolumn,showpacs,notitlepage]{revtex4-1}
\usepackage{multirow}
\usepackage{epsfig}
\usepackage{amsmath}
\usepackage{bm}
\usepackage{times}
\usepackage{graphicx}
\usepackage{color}
\usepackage{slashed}
\usepackage{graphicx}
\usepackage{amsmath} 
\usepackage[latin1]{inputenc}
\usepackage{hyperref}
\usepackage{soul}
\usepackage{multirow}
\usepackage{epsfig}
\usepackage{amsmath}
\usepackage{bm} 
\usepackage{times}
\usepackage{graphicx}
\usepackage{color}
\usepackage{slashed}
\usepackage{graphicx}
\usepackage{amsmath}

\def\bea{\begin{eqnarray}}
\def\eea{\end{eqnarray}}
\def\bean{\begin{eqnarray*}}
\def\eean{\end{eqnarray*}} 
\def\nn{\nonumber}
\def\beaal{\begin{align}}
\def\eeaal{\end{align}}

\begin{document}  
 
\title{Gravitational Wave Signatures of Gauged Baryon and Lepton Number}

\author{Jessica~Bosch\vspace{1mm}}
\affiliation{Department of Chemistry and Physics, Barry University, Miami Shores, Florida 33161, USA\vspace{1mm}}
\author{Zoraida~Delgado\vspace{1mm}}
\affiliation{Department of Chemistry and Physics, Barry University, Miami Shores, Florida 33161, USA\vspace{1mm}}
\author{Bartosz~Fornal}
\affiliation{Department of Chemistry and Physics, Barry University, Miami Shores, Florida 33161, USA\vspace{1mm}}
\author{Alejandra~Leon\vspace{1mm}}
\affiliation{Department of Chemistry and Physics, Barry University, Miami Shores, Florida 33161, USA\vspace{1mm}}

\date{\today}

\begin{abstract}

We demonstrate that novel types of gravitational wave signatures arise in theories with new gauge symmetries broken at high energy scales. For concreteness, we focus on models with gauged baryon number and lepton number, in which neutrino masses are generated via the type I seesaw mechanism, leptogenesis occurs through the decay of a heavy right-handed neutrino, and one of the new baryonic fields is a good dark matter candidate. Depending on the scalar content of the theory, the gravitational wave spectrum consists of contributions from cosmic strings, domain walls, and first order phase transitions. We show that  a characteristic double-peaked signal from domain walls or a sharp domain wall peak over a flat cosmic string background may be generated. Those new signatures are within the reach of future experiments, such as Cosmic Explorer, Einstein Telescope, DECIGO, Big Bang Observer, and LISA.
 \end{abstract}

\maketitle

\section{Introduction}
\label{intr}

In the Standard Model \cite{Glashow:1961tr,Higgs:1964pj,Englert:1964et,Weinberg:1967tq,Salam:1968rm,Fritzsch:1973pi,Gross:1973id,Politzer:1973fx}, baryon number  and lepton number  are accidental global symmetries of the  Lagrangian  of an unknown origin. Although one expects both of those symmetries to be violated if grand unification  is realized in Nature \cite{Georgi:1974sy,Fritzsch:1974nn}, so far no sign of related processes, such as proton decay, have been observed in experiments \cite{Super-Kamiokande:2016exg}, which excludes the minimal non-supersymmetric version of those theories. If grand unification does not happen, then  global baryon and lepton number may be a  low-energy manifestation of some more fundamental  gauge symmetries unbroken at high energy scales. Indeed, this line of reasoning is supported by the self-consistency of quantum theories of gravity, in which only  gauge symmetries can be properly accommodated, unless unnatural conditions are introduced \cite{Kallosh:1995hi}.

The first attempt to promote baryon and lepton number to the status of $\rm U(1)$ gauge symmetries dates back to the 1970s \cite{Pais:1973mi}, and was followed by further theoretical efforts  throughout the subsequent two decades \cite{Rajpoot:1987yg, Foot:1989ts, Carone:1995pu, Georgi:1996ei}. However, the first phenomenologically viable  model of this type was constructed only recently in \cite{FileviezPerez:2010gw}, and later modified to avoid all current experimental bounds \cite{Duerr:2013dza,Perez:2014qfa}. The idea of gauging baryon and lepton number was later successfully incorporated into a supersymmetric framework \cite{Arnold:2013qja}, theories unifying baryon number and color \cite{Fornal:2015boa,Fornal:2015one}, and generalized to a non-Abelian gauged lepton number  \cite{Fornal:2017owa}.
Models with  gauged baryon and lepton numbers have very attractive features: they explain the stability of the proton, have a natural realization of the seesaw mechanism for neutrino masses, contain an attractive baryonic dark matter candidate \cite{Duerr:2014wra,Ohmer:2015lxa,FileviezPerez:2018jmr}, and can accommodate high scale leptogenesis \cite{FileviezPerez:2021hbc}. Thus far, in all of the existing ${\rm U}(1)$ formulations of  theories with gauged baryon and  lepton number, each of the symmetries was broken by the vacuum expectation value of a single scalar. However, there is no reason to expect that the scalar sector is this minimal.

To this end, in this paper we investigate ways to probe the composition of high-scale symmetry breaking sectors, i.e., when at least one of the two sectors consists of more than one scalar breaking the symmetry. Although we focus on the class of theories with gauged baryon and lepton number, most of our analysis is  general and can be applied to other theories with two broken ${\rm U}(1)$ gauge symmetries. Conventional particle physics experiments are not able to  differentiate between the two scenario, or even probe them at all if the symmetry breaking scale is  high.  Nevertheless, as we demonstrate below, gravitational wave detectors have opened up a completely  new set of opportunities to probe such models.

A renaissance period for gravitational wave physics was initiated by the first direct detection  of a gravitational wave signal coming from a black hole merger by the Laser Interferometer Gravitational Wave
Observatory (LIGO) within the LIGO/Virgo collaboration \cite{LIGOScientific:2016aoc}. By now, over one hundred of such events, involving also neutron stars, have been recorded. Those discoveries provide an ideal opportunity to test general relativity, but they are not directly related to particle physics. The gravitational waves which enable probing particle physics models, although not yet discovered, are expected  to come in the form of a stochastic gravitational wave background produced in the early Universe by phenomena such as inflation \cite{Turner:1996ck}, first order phase transitions \cite{Kosowsky:1991ua}, domain walls \cite{Hiramatsu:2010yz} and cosmic strings \cite{Vachaspati:1984gt,Sakellariadou:1990ne}. Although for such signals to be detectable at LIGO the underlying particle physics models require a large  fine-tuning of parameters, future gravitational wave experiments, such as   the
Laser Interferometer Space Antenna (LISA) \cite{Audley:2017drz}, Cosmic Explorer \cite{Reitze:2019iox}, Einstein Telescope \cite{Punturo:2010zz}, DECIGO \cite{Kawamura:2011zz}, and Big Bang Observer \cite{Crowder:2005nr}, will be sensitive to more generic scenarios.

The most model-independent stochastic gravitational wave background comes from cosmic strings, which are topological defects  formed via the Kibble mechanism \cite{Kibble:1976sj} upon a spontaneous breaking of a ${\rm U}(1)$ symmetry. They correspond to one-dimensional field configurations along the direction in which the symmetry remains unbroken.  The dynamics of the  produced cosmic string network 
 provides a long-lasting source of gravitational radiation resulting in a mostly flat  stochastic gravitational wave background, with its strength dependent only on the scale of the ${\rm U}(1)$ breaking. Cosmic string signatures have been considered in the context of grand unified theories \cite{Buchmuller:2019gfy,King:2020hyd}, neutrino seesaw models \cite{Blanco-Pillado:2017oxo,Ringeval:2017eww,Cui:2017ufi,Cui:2018rwi,Guedes:2018afo,Dror:2019syi,Zhou:2020ils}, new physics at the high scale \cite{Gouttenoire:2019rtn}, as well as baryon and lepton number violation \cite{Fornal:2020esl}. For a review of gravitational waves signatures of  cosmic strings see \cite{Gouttenoire:2019kij}, and for the  constraints from LIGO/Virgo data see \cite{LIGOScientific:2021nrg}. 

The other topological defects which can be produced in the early Universe are domain walls, created when a $Z_2$ symmetry is spontaneously broken. They are two-dimensional field configurations existing at the boundaries of  regions corresponding to different vacua. In order for domain walls not to overclose the Universe, they need to annihilate away. This is possible when there exists a small energy density difference between the two vacua (the so-called potential bias). Domain wall annihilation leads to a stochastic gravitational wave background which is peaked at some frequency, but its strength and the peak frequency are, as in the case of cosmic strings, independent of the exact particle physics details of the model -- the spectrum is determined by only two  parameters: the scale of the symmetry breaking and the potential bias. Domain wall signatures have been considered in many theoreies beyond the Standard Model, including new  electroweak scale physics  \cite{Eto:2018hhg,Eto:2018tnk,Chen:2020soj,Battye:2020jeu}, supersymmetry  \cite{Kadota:2015dza},  axions \cite{Craig:2020bnv,Blasi:2022ayo}, grand unification \cite{Dunsky:2021tih}, models with left-right symmetry \cite{Borah:2022wdy}, baryon/lepton number violation \cite{Fornal:2023hri}, and models of leptogenesis  \cite{Barman:2022yos}. 
The physics of domain walls and the expected gravitational wave spectrum are reviewed in \cite{Saikawa:2017hiv}; the bounds on domain walls from LIGO/Virgo data can be found in \cite{Jiang:2022svq}.

The most model-dependent gravitational wave signatures arise from cosmological first order phase transitions. Those occur when the effective potential develops a new minimum with a lower energy density than the high-temperature one. If there exists a potential barrier between the two minima, the transition is first order and bubbles of true vacuum are being nucleated in various points in space. Such bubbles of true vacuum expand, eventually filling up the entire Universe. Gravitational waves are emitted from bubble collisions, turbulence, and sound shock waves in the primordial plasma generated by the violent expansion of the bubbles. The position of the gravitational wave peak is highly dependent on the temperature at which bubble nucleation occurs. First order phase transitions have been analyzed in a plethora of particle physics models, including, again, electroweak scale new physics \cite{Grojean:2006bp,Vaskonen:2016yiu,Dorsch:2016nrg,Bernon:2017jgv,Chala:2018ari,Angelescu:2018dkk,Alves:2018jsw,Han:2020ekm,Benincasa:2022elt}, 
supersymmetry \cite{Craig:2020jfv,Fornal:2021ovz},
 axions \cite{Dev:2019njv,VonHarling:2019rgb,DelleRose:2019pgi}, grand unification \cite{Croon:2018kqn,Huang:2020bbe,Okada:2020vvb}, baryon/lepton   number violation \cite{Hasegawa:2019amx,Fornal:2020esl}), neutrino seesaw models  \cite{Brdar:2018num,Okada:2018xdh,DiBari:2021dri,Zhou:2022mlz},  new flavor physics \cite{Greljo:2019xan,Fornal:2020ngq}, 
dark gauge groups \cite{Schwaller:2015tja,Breitbach:2018ddu,Croon:2018erz,Hall:2019ank}, models with conformal invariance \cite{Ellis:2020nnr,Kawana:2022fum}, and dark matter \cite{Baldes:2017rcu,Azatov:2021ifm,Costa:2022oaa,Costa:2022lpy,Fornal:2022qim,Kierkla:2022odc,Azatov:2022tii}.
A comprehensive review of gravitational waves  from first order phase transitions can be found in \cite{Caldwell:2022qsj,Athron:2023xlk}, while the most recent constraints from LIGO/Virgo data were derived in \cite{LIGO_FOPT}. For recent progress on supercooled phase transitions see \cite{Ellis:2019oqb,Lewicki:2019gmv,Lewicki:2020jiv,Ellis:2020nnr}.

In this work, we examine how gravitational wave signals from domain walls, cosmic strings, and  phase transitions interplay with each other, producing novel features in the expected spectrum. The two new gravitational wave signatures which have not been considered in the literature so far are:\break $(1)$ Two coexisting signals from domain wall annihilation, forming a characteristic sharp double-peak in the spectrum; $(2)$ Domain wall signal over a flat cosmic string contribution, leading to an unusually-shaped peak. 
Although  we focus on a specific model, our results involving cosmic strings and domain walls, in particular signatures $(1)$ and $(2)$, are general, since they do not depend on the details of the model.

\section{The model}\label{modell}

The gravitational wave signatures we propose to search for, as will be discussed in Section \ref{signatures}, are anticipated in a large class of models with a two-step symmetry breaking pattern. In this paper, to provide a concrete realization of such scenarios, we focus on a model with gauged baryon and lepton number, based on the gauge group
\bea\label{group}
{\rm SU}(3)_c \times {\rm SU}(2)_L \times {\rm U}(1)_Y \times {\rm U}(1)_B \times {\rm U}(1)_L  \ .
\eea
Below we describe the possible symmetry breaking patterns in the model and the new particles along with their masses.

\subsection*{Scalar sector}

As mentioned in Section \ref{intr}, we consider  extending the usual single-scalar symmetry breaking sector to possibly include two scalars per each symmetry breaking. 
Each of the fields breaking the ${\rm U}(1)_L$  symmetry comes in the  representation
\bea
\Phi_{Li} = (1,1,0,0,2) \ ,
\eea
while each of the scalars breaking  ${\rm U}(1)_B$ is
\bea
\Phi_{Bi} = (1,1,0,3,3) \ .
\eea
Within this framework, there  are four possible cases:
\begin{itemize}
\item[$(a)$] $\Phi_B$  breaks ${\rm U(1)}_B$ and   $\Phi_L$   breaks ${\rm U(1)}_L$,
\item[$(b)$] $\Phi_{B1}$, $\Phi_{B2}$  break ${\rm U(1)}_B$ and  $\Phi_L$   breaks ${\rm U(1)}_L$,
\item[$(c)$] $\Phi_{B}$ breaks ${\rm U(1)}_B$ and  $\Phi_{L1}$, $\Phi_{L2}$   break ${\rm U(1)}_L$,
\item[$(d)$] $\Phi_{B1}$, $\Phi_{B2}$  break ${\rm U(1)}_B$ and  $\Phi_{L1}$, $\Phi_{L2}$   break ${\rm U(1)}_L$.
\end{itemize}

We assume that the mixed terms involving scalars breaking different ${\rm U}(1)$ gauge groups have negligible coefficients. This implies that for a given ${\rm U}(1)$, if only one scalar breaks the symmetry, the scalar potential is
\bea
V(\Phi) &=& -m^2 |\Phi|^2  +  \lambda |\Phi|^4 \ ,
\eea
whereas if two scalars participate in symmetry breaking,
\bea\label{2hpot}
V(\Phi_1,\Phi_2) &=& -  \,m_1^2 |\Phi_1|^2  +  \lambda_{1} |\Phi_1|^4  -m_2^2 |\Phi_2|^2  + \lambda_{2} |\Phi_2|^4\nn\\
&+& \left[(\lambda_4|\Phi_1|^2 \!+\! \lambda_5 |\Phi_2|^2 \!+\! \lambda_6 \Phi_1^* \Phi_2)\Phi_1^*\Phi_2 + {\rm h.c.}\right]\nn\\
&-&(m_{12}^2 \Phi_1^* \Phi_2 + {\rm h.c.})  + \lambda_3 |\Phi_1|^2 |\Phi_2|^2 \ .
\eea
The scalars develop the following vacuum expectation values,
\bea
\langle \Phi_i \rangle = 
\frac{v_i}{\sqrt2} \ .
\eea
Whenever two scalars take part in the symmetry breaking, we define
$
v \equiv \sqrt{v_1^2 + v_2^2} \ .
$
This way one can collectively describe the ${\rm U}(1)_B$ breaking scale as  $v = v_B$, and the ${\rm U}(1)_L$ breaking scale as  $v = v_L$, independent of whether the vacuum expectation value comes from a single scalar or two scalars.

If lepton number is broken at a higher scale than baryon number, the symmetry breaking pattern is:
 \begin{equation}
     \begin{array}{c}
    {\rm SU}(3)_c \times {\rm SU}(2)_L \times {\rm U}(1)_Y \times {\rm U}(1)_B \times {\rm U}(1)_L \nn\\[3pt]
\hspace{11.2mm}\bigg\downarrow \hspace{3mm}{\scriptstyle \langle\Phi_{Li}\rangle \ \ne \ 0} \nn\\[12pt]
{\rm SU}(3)_c \times {\rm SU}(2)_L \times {\rm U}(1)_Y \times {\rm U}(1)_B \nn\\[3pt]
\hspace{11.2mm}\bigg\downarrow \hspace{3mm}{\scriptstyle  \langle\Phi_{Bi}\rangle \ \ne \ 0} \nn\\[12pt]
 {\rm SU}(3)_c \times {\rm SU}(2)_L \times {\rm U}(1)_Y \ ,\\[7pt]
     \end{array}
   \end{equation}  
followed by the usual electroweak symmetry breaking  by the Standard Model Higgs. We note that the order of ${\rm U}(1)_B$ and ${\rm U}(1)_L$ breaking may be reversed.

\subsection*{Fermion sector}

To provide a concrete quantitative example, we consider the model with gauged ${\rm U}(1)_B$ and ${\rm U}(1)_L$ proposed in \cite{Perez:2014qfa}, which involves the minimal fermionic particle content for a theory with such a gauge group. It is straightforward to check that all  gauge anomalies are cancelled if the Standard Model quark fields $Q_L^j$, $u_R^j$, $d_R^j$ and lepton fields $l_L^j$, $e_R^j$ are augmented by
\bea\label{fermions}
\nu_{R}^j &=& (1,1,0,0,1) \ , \nn\\
\Psi_L &=& \begin{pmatrix}
\psi_L^+ \\[2pt]
\psi_L^0
\end{pmatrix} = \left(1,2,\tfrac12,\tfrac32,\tfrac32\right) \ , \nn\\
\Psi_R &=& \begin{pmatrix}
\psi_R^0 \\
\psi_R^-
\end{pmatrix} = \left(1,2,\tfrac12,-\tfrac32,-\tfrac32\right) \ , \nn\\
\Sigma_L &=& \tfrac12 \begin{pmatrix}
\sigma^0 & \sqrt2 \,\sigma^+ \\[2pt]
\sqrt2 \,\sigma^- & - \sigma^0
\end{pmatrix} =  \left(1,3,0,-\tfrac32,-\tfrac32\right) \ , \nn\\
\chi_L &=&  \left(1,1,0,-\tfrac32,-\tfrac32\right) \ ,
\eea
where $j$ is the family index.
Among the fields above, $\nu_R^j$ are the right-handed neutrinos, whereas $\chi_L$ is a Majorana dark matter candidate discussed in Section \ref{dam}.

\subsection*{Particle masses}

The scalars $\Phi_{Bi}$ and $\Phi_{Li}$ generate masses for the new fermions through the following Lagrangian terms,
\bea\label{9}
-\mathcal{L}  &\supset&  \sum_i \!\left(Y_{\Psi}^i \overline\Psi_R\Psi_L \Phi_{Bi}^* \!+\! Y_{\Sigma}^i {\rm Tr}(\Sigma_L^2) \Phi_{Bi} \!+\! Y_{\chi}^i \chi_L \chi_L \Phi_{B i} \right)\nn\\
&+&  y_\nu^j \,\bar{l}_L^j H \nu_R^{j} + \sum_i Y_{\nu}^{ij} \nu_R^j \nu_R^j \Phi_{Li} + {\rm h.c.} \ ,  \ \ \ \ 
\eea
which provide  vector-like masses to the new fermions, as well as  introduce a type I seesaw mechanism for the neutrinos. For example, assuming that the Yukawa couplings are $y_\nu^j \sim 1$ and $Y_{\nu}^j\sim 10^{-2}$, the measured neutrino mass splittings are reproduced if $v_L \sim 10^5 \ {\rm PeV}$. The mass matrices for the new fermions are provided in \cite{Perez:2014qfa}.

The spontaneous breaking of ${\rm U}(1)_L$ and ${\rm U}(1)_B$ leads to the appearance of vector gauge bosons $Z_L$ and $Z_B$. Given the charges of the scalars breaking the two symmetries, the corresponding masses are 
\bea
m_{Z_L} = 2 g_L v_L \ , \ \ \ \ \ \ \ \  m_{Z_B} = 3 g_B v_B \ , 
\eea
where $g_L$ and $g_B$ are the ${\rm U}(1)_L$ and ${\rm U}(1)_B$ gauge couplings, respectively, whose values are free parameters.

\section{Dark matter and matter-antimatter asymmetry}
\label{dam}

Apart from providing a natural framework accommodating a type I  seesaw mechanism generating small neutrino masses via ${\rm U}(1)_L$ breaking, the model also contains  a phenomenologically viable dark matter candidate $\chi_L$ \cite{Perez:2014qfa,Ohmer:2015lxa} and can account for the matter-antimatter asymmetry through leptogenesis \cite{FileviezPerez:2021hbc}. We discuss the most relevant aspects  of those highlights of the model below.

\subsection*{Dark matter}

After ${\rm U}(1)_B$ breaking, there remains a residual discrete ${Z}_2$ symmetry under which the new fermions transform as
\bea
\Psi_L &\to& - \Psi_L \ , \ \ \ \ \overline\Psi_R \to - \overline\Psi_R \ ,\nn\\
\Sigma &\to& -\Sigma \ , \ \ \ \ \ \ \ \chi_L \to -\chi_L \ .
\eea
 If $\chi_L$ is the lightest of the new fermions, there is no decay channel available for it, thus it becomes a good candidate for particle dark matter.
 
 It was argued in   \cite{Duerr:2014wra,FileviezPerez:2018jmr} that in models with gauged baryon and lepton number consistency with the dark matter relic abundance of $h^2 \Omega_{\rm DM} \approx 0.12$ \cite{Planck:2018vyg} imposes an upper bound on the ${\rm U(1)}_B$ breaking scale. In particular, if the dark matter annihilation happens via the resonant $s$-channel process
 \bea
 \chi_L\,\chi_L \to Z_B^* \to \bar{q} \,q \ ,
 \eea
a dependence between the parameters $v_B$, $Y_\chi$, and $g_B$ arises, and the perturbativity requirement  leads to  $g_B v_B \lesssim 20 \ {\rm TeV}$. 
 This was the reason why in \cite{Fornal:2020esl}, where the gravitational wave signal from a model with gauged baryon and lepton number was considered, a low scale of  ${\rm U}(1)_B$ was imposed.

 However, as was demonstrated  in \cite{Ohmer:2015lxa}, in the  model we are considering other dark matter annihilation channels remain unsuppressed, including the nonresonant $t$-channel  process 
  \bea
 \chi_L\,\chi_L \to \Phi_B\, \Phi_B \ ,
 \eea
 whose cross section can be sufficiently large to explain the dark matter relic density. Therefore, the arguments   in \cite{Duerr:2014wra,FileviezPerez:2018jmr} do not apply in our case, and the scale of ${\rm U}(1)_B$ breaking can be high. Alternatively, the aforementioned bound on the ${\rm U}(1)_B$ breaking scale can always  be avoided by assuming  nonthermal dark matter production.

\subsection*{Leptogenesis}

There cannot exist any primordial baryon or lepton number asymmetry
above the scales of ${\rm U}(1)_B$ and ${\rm U}(1)_L$ breaking. An excess of matter over antimatter can only arise once one of those two symmetries is broken. A natural setting to achieve this below the scale of ${\rm U}(1)_L$ breaking  is offered by high-scale leptogenesis (see \cite{Davidson:2008bu} and references therein), in which  a lepton number asymmetry is generated through the out-of-equilibrium decays of the lightest right-handed neutrino,
\bea\label{ndecay}
N_1 \to H\, l_L \ .
\eea
The $CP$ asymmetry is introduced through the standard interference between the tree-level diagram for the process in Eq.\,(\ref{ndecay}) and the one-loop diagrams involving $H$, $l_L$, and the two heavier right-handed neutrinos $N_2$, $N_3$ in the loop. 

The generated lepton asymmetry is calculated by solving the Boltzmann equations for the evolution of the lightest right-handed neutrino abundance $Y_{N_1}= n_{N_1}/s$ (where $n_{N_1}$ is the $N_1$ particle density  and $s$ is the co-moving entropy density) and the $B\!-\!L$ asymmetry $Y_{B-L}$ \cite{Buchmuller:2004nz},
\bea
\frac{d Y_{N_1}}{d z} &=&  -(D+S)(Y_{N_1} - Y_{N_1}^{\rm eq})  \ ,\nn\\
\frac{d Y_{B-L}}{d z} &=&  -\epsilon_1 D (Y_{N_1} - Y_{N_1}^{\rm eq}) - W Y_{B-L}  \ ,
\eea
where $z = m_{N_1}/T$, the term $D$ accounts for decays and inverse decays, $S$ represents $\Delta L=1$ scatterings, $W$ describes the washout effects, and $\epsilon_1$ is the $CP$ asymmetry parameter. 
In our case, the Boltzmann equations are slightly different than in the standard leptogenesis scenario, since   the right-handed neutrinos have an extra interaction with the $Z_L$ gauge boson. Those equations were solved in \cite{FileviezPerez:2021hbc}, and the amount of  the generated lepton asymmetry $\Delta L$ was determined for various values of model parameters.

The produced lepton asymmetry is then partially converted into a baryon asymmetry through the electroweak sphalerons. Above the scale of ${\rm U}(1)_B$ breaking the sphaleron-induced interactions have the form 
\bea
(QQQL)^3 \overline\Psi_R \Psi_L \Sigma_L^4 \ ,
\eea
and it was shown in  \cite{Duerr:2014wra} that if the breaking of ${\rm U}(1)_B$ occurs close to the electroweak scale, then the final baryon asymmetry predicted by the model is given by
\bea
|\Delta B| = \frac{32}{99} |\Delta L| \ .
\eea
This can explain the observed baryon-to-photon ratio  \cite{ParticleDataGroup:2022pth}
\bea
\eta \approx  6 \times 10^{-10} 
\eea
if  the scale of ${\rm U}(1)_L$ breaking satisfies the relation
\bea\label{lepbound}
v_L \gtrsim 4000 \ {\rm PeV} \ .
\eea
The requirement for such a high ${\rm U}(1)_L$ symmetry breaking scale provides the desired setting accommodating  the type I seesaw mechanism for the neutrinos.

\section{Cosmic string spectrum}
\label{CSsec}

Spontaneous breaking of a ${\rm U}(1)$ gauge symmetry leads to the production of topological defects in the form of cosmic strings \cite{Kibble:1976sj}, which correspond to one-dimensional field configurations along the direction of the unbroken symmetry. 
The network of produced cosmic strings is described collectively  by the string tension $\mu$, equal to the energy stored in a string per unit length, and depends solely on the scale at which the ${\rm U}(1)$ gauge symmetry is broken  \cite{Vilenkin:2000jqa,Gouttenoire:2019kij},
\bea
G\mu =2\pi \left(\frac{v}{M_{Pl}}\right)^2 ,
\eea
where $M_{Pl}=1.22\times 10^{13} \ {\rm PeV}$ is the Planck mass, the gravitational constant $G = 6.7\times 10^{-39} \ \rm GeV^{-2}$, and the winding number was taken to be one. The constraints from the cosmic microwave background measurements set an upper limit on the string tension of  $G\mu\lesssim 10^{-7}$ \cite{Ade:2013xla}, which corresponds to the following bound on the scale of symmetry breaking,
\bea\label{cmbv}
v \lesssim 1.5 \times 10^{9} \ {\rm PeV} \ .
\eea
 Through its dynamics, the cosmic string network 
 provides a long-lasting source of gravitational radiation and leads to a stochastic gravitational wave background roughly constant across a wide range of frequencies.

\subsection*{Dynamics of cosmic strings}

There are two main processes governing the behavior of a cosmic string network: formation of string loops and stretching due to the expansion of the Universe. The first of those contributions, creation of string loops, happens  when   long strings intersect and intercommute. The newly created string loops oscillate and emit gravitational radiation,  mainly from cusps and kinks propagating through the string loop, and from kink-kink collisions \cite{Olum:1999sg, Moore:2001px}.

A competition between these two effects leads to the so-called scaling regime, in which there is a large number of string loops and a small number of Hubble-size strings \cite{Kibble:1984hp, PhysRevLett.60.257, PhysRevLett.63.2776, PhysRevD.40.973, PhysRevLett.64.119}. There is a continuous flow of energy from long strings to string loops, and then to gravitational radiation through their decays. This gravitational radiation makes up a fixed fraction of the energy density of the Universe  \cite{Hindmarsh:1994re}.

To describe this process quantitatively, we follow the steps outlined in   \cite{Cui:2018rwi,Gouttenoire:2019kij}. We consider a string loop created at time $t_0$  with initial  length $l(t_0)=\alpha\,t_0$, where $\alpha$ is a constant loop size parameter.  The loop oscillates emitting  gravitational waves with frequencies 
\bea
\tilde f = \frac{2k}{l}  \ , 
\eea
where $k$ is a positive integer. Rescaling this result by the scale factor $a(t)$, one obtains the currently observed frequency,
\bea
f = \frac{a(t_e)}{a(T)} \,\tilde f \ ,
\eea
where $t_e$ is the time of the emission and $T$ is the time today. 
The spectrum of the emitted gravitational waves from a single oscillating string loop is  given by \cite{Blanco-Pillado:2013qja,Blanco-Pillado:2017oxo}
\bea
P_{(k,n)} = \frac{\Gamma G\mu^2}{k^n} \bigg(\sum_{p=1}^\infty \frac1{p^n}\!\bigg)^{-1},
\eea
where for the contribution from cusps $n=4/3$, from kinks $n=5/3$, and from kink-kink collisions $n=2$, while the overall factor $\Gamma \simeq 50$ \cite{PhysRevD.31.3052}.
Due to the constant emission of gravitational radiation, the string loop shrinks and its length at the time of the gravitational wave emission is
\bea
l(t_e) = \alpha\,t_0 - \Gamma G\mu \,(t_e-t_0) \ ,
\eea
causing the  loop to vanish after the time ${\alpha\,t_0}/{(\Gamma G \mu)}$.

The only model-dependent quantity describing the  cosmic string network is the  loop distribution function $F(l,t_0)$ for the created loops. Adopting the well-established model developed in \cite{Martins:1995tg,Martins:1996jp,Martins:2000cs}, describing the string network just by the  mean string velocity and the correlation length, leads  in the scaling regime to the following formula,
\bea
F(l,t_0) = \frac{\sqrt2 \,{C}_{\rm eff}}{\alpha\,t_0^4}\,\delta(l-\alpha\,t_0) \ ,
\eea
where the constant ${C}_{\rm eff}$ depends on the era in the evolution of the Universe (for radiation $C_{\rm eff} = 5.4$, whereas for the matter dominated era $C_{\rm eff} = 0.39$  \cite{Cui:2018rwi}).

\subsection*{Gravitational wave spectrum}

The stochastic gravitational wave background generated by the dynamics of the cosmic string network  is  \cite{Cui:2018rwi,Gouttenoire:2019kij}
\bea\label{CSspectr}
h^2\Omega_{\rm CS}(f) &=&\frac{2 h^2\mathcal{F}_\alpha}{\rho_c\alpha^2 f}\,\sum_{k,n}\,{k\, P_{(k,n)}} \int_{t_F}^{T}\!dt_e\ \frac{C_\text{eff}(t_{0k})}{t_{0k}^{\, 4}}\ \ \ \ \nn\\
&\times&\left(\frac{a(t_e)}{a(T)}\right)^{\!5}\left(\frac{a(t_{0k})}{a(t_e)}\right)^{\!3}\theta(t_{0k}-t_{F}) \ ,
\eea
where
\bea\label{times}
t_{0k}=\frac{1}{\alpha}\left(\frac{2k}{f}\frac{a(t_e)}{a(T)}+\Gamma G\mu \, t_e\right) \ .
\eea
In Eqs.\,(\ref{CSspectr}) and (\ref{times}) the parameters are the following: $\mathcal{F}_\alpha$ is the fraction of the loops contributing to the gravitational wave signal, estimated to be  $\mathcal{F}_\alpha \approx 0.1$  \cite{Blanco-Pillado:2013qja} since the majority of the energy is lost by long strings going into highly boosted smaller loops which provide only a  subdominant contribution; $\rho_c$ is the critical density of the Universe; the loop size parameter $\alpha=0.1$  provides an accurate estimate of the loop size distribution \cite{Blanco-Pillado:2013qja,Blanco-Pillado:2017oxo}; $t_F$ is the time at which the cosmic string network  was formed, related to the density of the Universe at that time via $\sqrt{\rho(t_F)}=\mu$ \cite{Gouttenoire:2019kij}, $t_{0k}$ is the instance when the loop was  produced, and $\theta(x)$ is the Heaviside step function. We also note that, as argued in \cite{Cui:2018rwi,Gouttenoire:2019kij}, the largest contribution to the gravitational wave signal comes from the cusps.

\begin{figure}[t!]
\includegraphics[trim={2.4cm 0.8cm 1cm 0cm},clip,width=9.5cm]{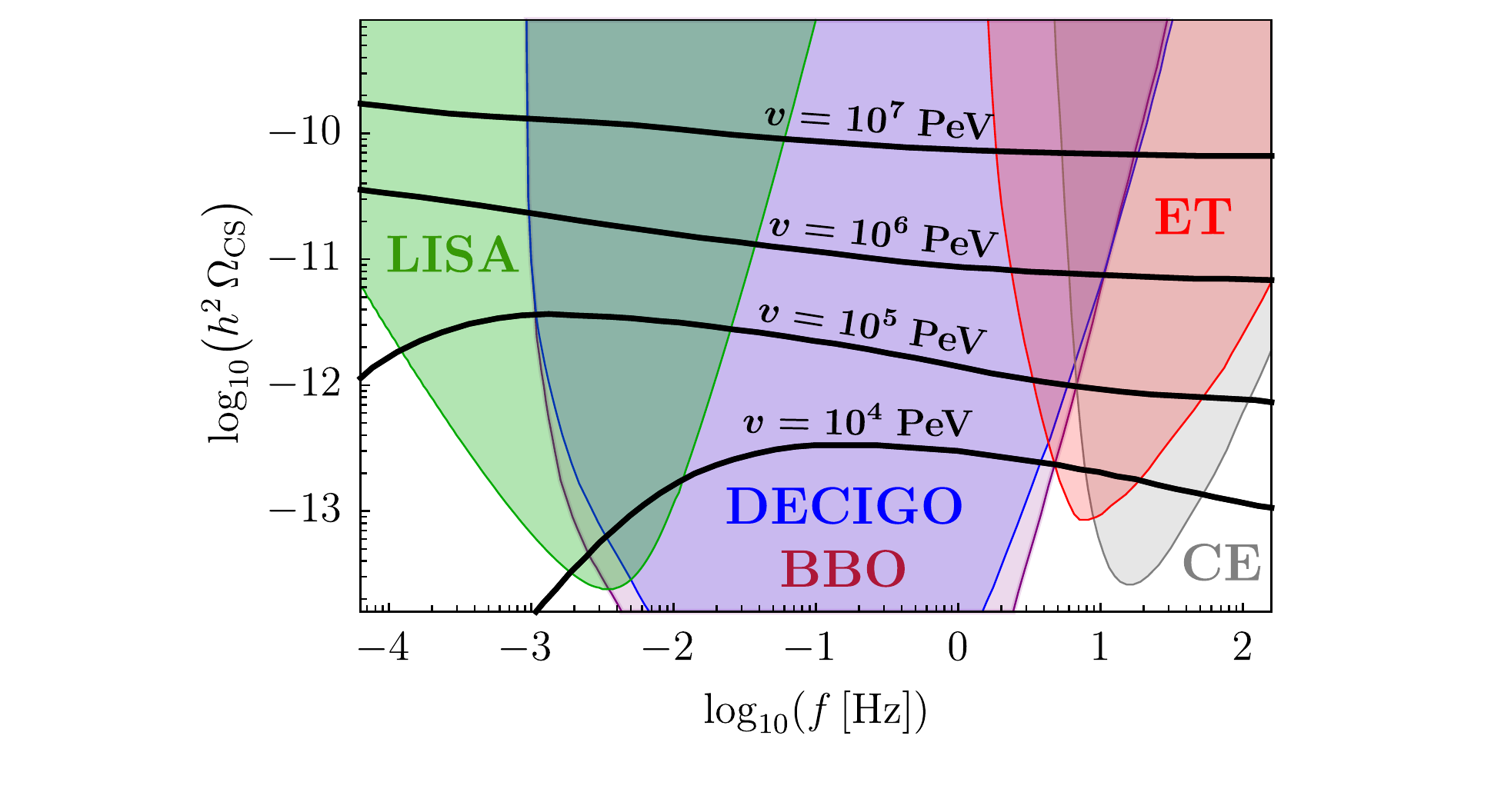} \vspace{-6mm}
\caption{Stochastic gravitational wave background from cosmic strings for four different symmetry breaking scales. Shaded regions correspond to the sensitivity of future gravitational wave detectors: LISA (green), DECIGO (blue), Big Bang Observer (purple), Einstein Telescope (red), and Cosmic Explorer (gray).}\label{fig:1}
\end{figure}

\begin{figure}[t!]
\includegraphics[trim={2.4cm 0.8cm 1cm 0cm},clip,width=8.8cm]{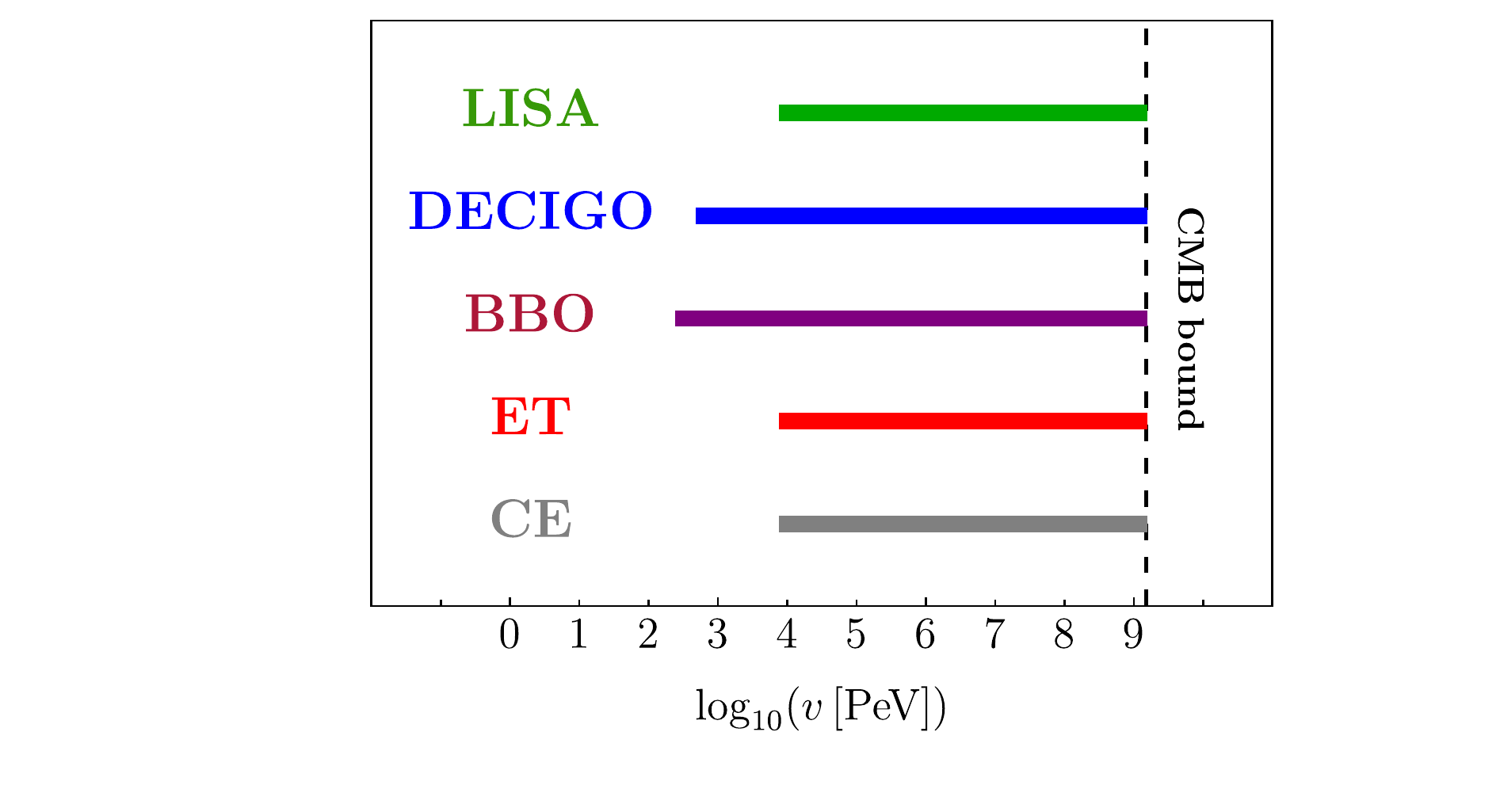} \vspace{-6mm}
\caption{Reach of future detectors in probing the scale of a ${\rm U}(1)$ symmetry breaking leading to the production of cosmic strings. The colors for each experiment correspond to those adopted in Fig.\,\ref{fig:1}.}\label{fig:2}
\end{figure}

The resulting stochastic gravitational wave background  is presented in Fig.\,\ref{fig:1} for four values of the symmetry breaking scale, in the range of frequencies relevant for the upcoming gravitational wave experiments, whose sensitivities are denoted by the colored regions. If $v \gtrsim 10^4 \ {\rm PeV}$, the signal can be seen by all the detectors: LISA \cite{Audley:2017drz}, DECIGO \cite{Kawamura:2011zz}, Big Bang Observer \cite{Crowder:2005nr},  Einstein Telescope \cite{Punturo:2010zz}, and Cosmic Explorer \cite{Reitze:2019iox}. This is illustrated in more detail in Fig.\,\ref{fig:2}, which shows the reach of each experiment in terms of the symmetry breaking scale leading to a cosmic string signal. The lower bound is detector-specific, whereas the upper bound reflects the cosmic microwave background constraint in Eq.\,(\ref{cmbv}).

The   cosmic string network  will be produced if either ${\rm U}(1)_B$ or ${\rm U}(1)_L$ is spontaneously broken by a single scalar. Therefore, among the cases enumerated in Section\,\ref{modell}, the  gravitational wave signals discussed here  are relevant  in case $(a)$ for both baryon and lepton number breaking, in case $(b)$ only for lepton number breaking, and in case $(c)$ only for baryon number breaking.

\section{Domain wall spectrum}
\label{DWsec}

Another kind of topological defects, appearing when a ${\rm U}(1)$ symmetry is broken by two scalars, are domain walls. As in the case of ${\rm SU}(2)$ breaking discussed in \cite{Fornal:2022qim}, the effective potential $V(\phi_1,\phi_2, T)$, which at high temperature has just one vacuum at $(\phi_1,\phi_2)=(0,0)$,  at lower temperature develops four vacua. They come in two pairs  related  via a gauge transformation, $\Phi_i \to e^{i\theta} \Phi_i$, and only two of them, say $\vec{\phi}_{\rm vac1}$ and $\vec{\phi}_{\rm vac2}$, correspond to disconnected manifolds, and thus are physically distinct \cite{Ginzburg:2004vp,Battye:2011jj}. When a transition takes place, patches of the Universe end up in either one of those vacua, leading to the creation of domain walls, i.e., two-dimensional field configurations  on the boundaries of $\vec{\phi}_{\rm vac1}$ and $\vec{\phi}_{\rm vac2}$.

If the two vacua  have identical energy densities, domain walls  remain stable and considerably affect the evolution of the Universe, introducing unacceptably 
large density fluctuations \cite{Saikawa:2017hiv}. Therefore, for a phenomenologically viable scenario, the ${Z}_2$ symmetry between the two vacua needs to be softly broken. It cannot be strongly broken, since then patches of the Universe would transition preferentially to the lower energy density vacuum and domain walls would not form. 
In our case, the soft breaking of the ${Z}_2$ symmetry removing the degeneracy between the vacua is provided by the terms involving $m_{12}^2$, $\lambda_4$ and $\lambda_5$ in the Lagrangian in Eq.\,(\ref{2hpot}).

\subsection*{Dynamics of domain walls}

The profile of the domain wall configuration $\vec\phi_{dw}(z)$ is the solution to the equation  \cite{Chen:2020soj}
\bea
\frac{d^2 \vec\phi_{dw}(z)}{d z^2} - \vec\nabla_{\!\phi}V_{\rm eff}\big[\vec\phi_{dw}(z)\big]= 0 \ ,
\eea
where the $z$-axis was chosen to be perpendicular to the domain wall, and the 
 boundary conditions are,
 \bea 
 \vec\phi_{dw}(-\infty) = \vec\phi_{\rm vac1} \ , \ \ \ \  \ \vec\phi_{dw}(\infty) = \vec\phi_{\rm vac2}.
 \eea

 As mentioned earlier, there are only two parameters which describe the gravitational wave spectrum from domain walls. 
 The first of them is the domain wall tension $\sigma$ given by
\bea
\sigma = \int_{-\infty}^\infty dz \!\left[ \frac12 \bigg(\frac{d\vec\phi_{dw}(z)}{d z}\bigg)^2 + V_{\rm eff}\big[\vec\phi_{dw}(z)\big]\right].
\eea
In the model we are considering, to a good approximation
\bea
\sigma \,\sim \,v^3 \ .
\eea
The other parameter is the potential bias $\Delta \rho$, i.e., the energy density difference between the vacua, in our case equal to
\bea\label{potbias}
\Delta \rho = \big(m_{12}^2+\tfrac12\lambda_4 v_1^2 + \tfrac12\lambda_5 v_2^2\big)\, v_1 v_2  \ .
\eea

The created domain walls are unstable and undergo annihilation, emitting gravitational radiation, provided that \cite{Saikawa:2017hiv}
\bea\label{con11}
\Delta \rho \gtrsim \left(\frac{\sigma}{M_{Pl}}\right)^2 \ .
\eea
If  in Eq.\,(\ref{potbias}) the term involving $m_{12}^2$ is the dominant one, then Eq.\,(\ref{con11}) takes the form $m_{12} \gtrsim {v^2}/{M_{Pl}}$. For example, if the symmetry breaking scale is $v \sim 10^3 \ \rm PeV$, this implies $m_{12} \gtrsim 100 \ {\rm MeV}$. An independent  constraint arises from the necessity of domain wall annihilation happening before Big Bang nucleosynthesis, so  the ratios of the produced elements are not altered, but the bound in Eq.\,(\ref{con11}) remains stronger.
\vspace{-1mm}

\subsection*{Gravitational wave spectrum}

\vspace{-1mm}
Domain wall annihilation leads to a
 stochastic gravitational wave background given by \cite{Kadota:2015dza,Saikawa:2017hiv},
 \vspace{-2mm}
\bea\label{dme}
h^2 \Omega_{\rm DW}(f)&\approx& 7.1\times 10^{-33}\left(\frac{\sigma}{\rm PeV^3}\right)^4\left(\frac{\rm TeV^4}{\Delta \rho}\right)^2\left(\frac{100}{g_*}\right)^{\frac13}\nn\\
&\times& \!\!\bigg[\!\left(\frac{f}{f_d}\right)^3\!\theta(f_d-f) + \left(\frac{f_d}{f}\right)\theta(f-f_d)\bigg], \ \ \ \ \ \ \ 
\eea
where for the area parameter we used  $\mathcal{A} = 0.8$ and for the efficiency parameter we adopted the value $\tilde\epsilon_{\rm gw} = 0.7$ \cite{Hiramatsu:2013qaa},\break
$\theta$ denotes the step function, and the peak frequency $f_d$ is 
\bea
f_d \approx (0.14 \ {\rm Hz}) \,  \sqrt{\frac{\rm PeV^3}{\sigma}\frac{\Delta \rho}{\rm TeV^4}} \ .
\eea
The slope of the signal falls $\sim f^3$ to the left of the peak when moving toward lower frequencies, and falls like $f^{-1}$ to the right of the peak when moving toward higher frequencies. The cosmic microwave background constraint on the strength of the signal at the peak is  $h^2 \Omega(f) < 2.9 \times 10^{-7}$ \cite{Clarke:2020bil}, which translates to the following condition on the  parameters,
\bea\label{cmmb}
\frac{\sigma}{\sqrt{\Delta\rho}} \lesssim 2.5\times 10^{12} \ \rm PeV \ ,
\eea
stronger than the bound imposed by Eq.\,(\ref{con11}).

\begin{figure}[t!]
\includegraphics[trim={2.4cm 0.8cm 1cm 0cm},clip,width=9.5cm]{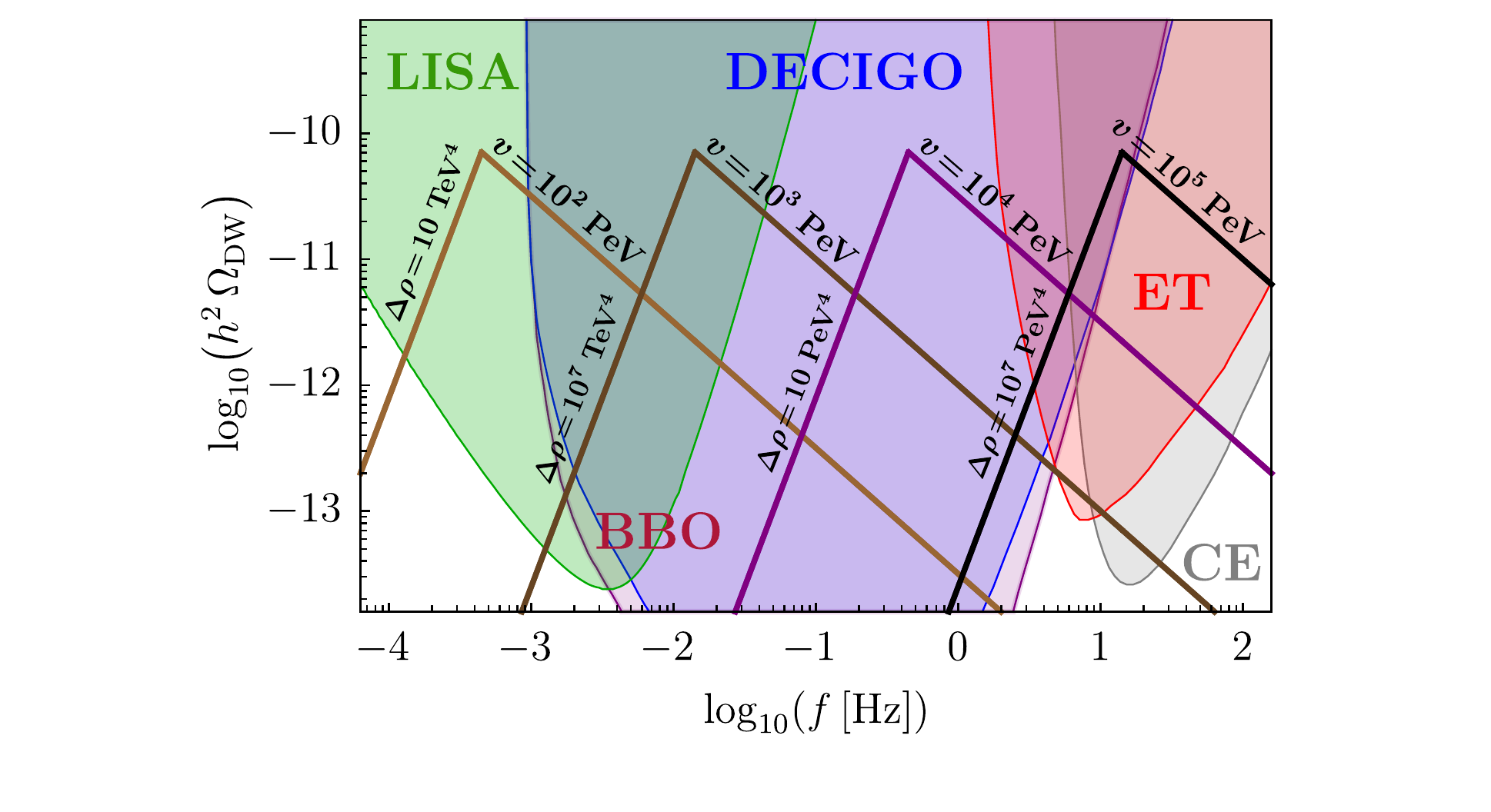} \vspace{-7mm}
\caption{Stochastic gravitational wave background from domain walls for various symmetry breaking scales. Shaded regions correspond to the sensitivity of future gravitational wave detectors, as in Fig.\,\ref{fig:1}.}\label{fig:3}
\end{figure}

\begin{figure}[t!]
\includegraphics[trim={2cm 0.8cm 1cm 0cm},clip,width=9.0cm]{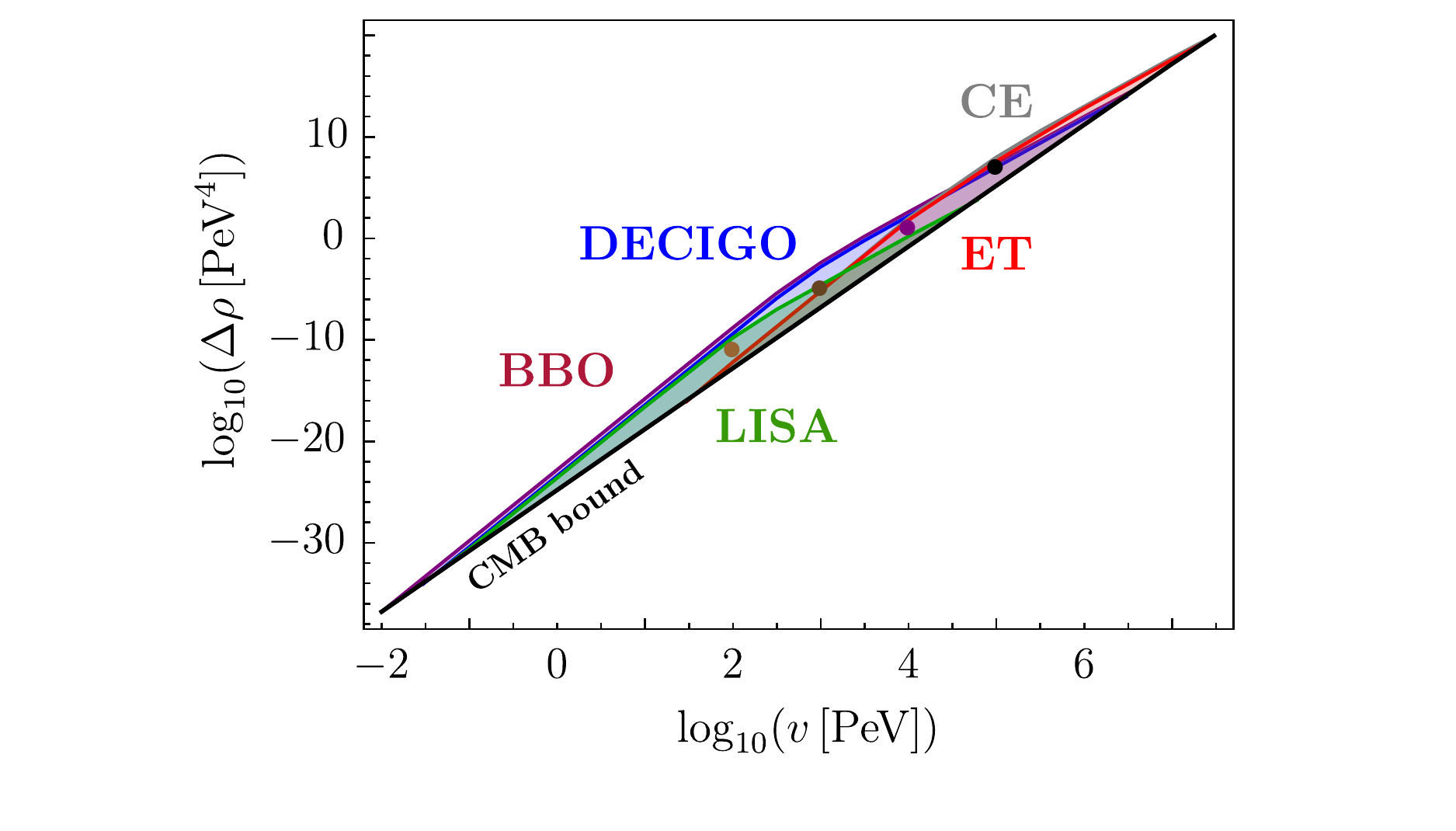} \vspace{-5mm}
\caption{Regions of parameter space $(v,\Delta\rho)$ for which the signal-to-noise ratio of the gravitational wave signal generated by domain wall annihilation is greater than five upon one year of data taking by various experiments. The choice of colors matches that in Fig.\,\ref{fig:3}, including the color of the dots which correspond to the four curves.}\label{fig:4}
\end{figure}

Several examples of gravitational wave spectra from domain wall annihilation, plotted using  Eq.\,(\ref{dme}), are shown in Fig.\,\ref{fig:3} for representative values of the parameters $v$ and $\Delta\rho$. The reach of the upcoming gravitational wave detectors is also shown, including LISA, Big Bang Observer, DECIGO, Einstein Telescope, and Cosmic Explorer. A more detailed look at their sensitivity is provided by Fig.\,\ref{fig:4}, which shows the full parameter space that can be probed by those experiments. The lower bound on the domain wall parameter $\Delta\rho$ is a reflection of the cosmic microwave background constraint from Eq.\,(\ref{cmmb}). The parameter $\Delta\rho$ depends in general on all three fundamental Lagrangian parameters $m_{12}$, $\lambda_4$, and $\lambda_5$ through the relation in Eq.\,(\ref{potbias}). Under the assumption that the term involving $m_{12}$ is dominant, the  experimental sensitivity plot in the plane $(v, m_{12})$ would be the same as in figure 4 of \cite{Fornal:2023hri}.

\section{First order phase transition spectrum}
\label{PTsec}

Perhaps the most anticipated stochastic gravitational wave signal to be discovered is the one generated by a first order phase transition in the early Universe,  predicted  in a large class of theories beyond the Standard Model. Such a signature in the case of ${\rm U}(1)_B$ breaking has been considered in \cite{Fornal:2020esl}, but in this work we adopt a different assumption for the Yukawa couplings and keep our analysis more general, so that it can be applied to both gauged ${\rm U}(1)_B$ and gauged ${\rm U}(1)_L$. A first order phase transition can occur either when the symmetry breaking sector  consists of a single scalar, or contains multiple scalars. Since the generalization is straightforward, we concentrate on the case with a single scalar.

\subsection*{Effective potential}

The effective potential for the background field $\phi$ consists of the tree-level part, the one-loop Coleman-Weinberg zero temperature correction, and the finite temperature contribution. Upon imposing the condition that the minimum of the zero temperature potential and the mass of $\phi$ remain at their tree-level values (i.e., the cutoff regularization scheme),  the effective potential is given by
\bea
&&V_{\rm eff}(\phi, T) = -\frac12 \lambda v^2 \phi^2 + \frac14\lambda \phi^4 \nn\\
&&+  \sum_i \frac{n_i m_i^2(\phi)}{64\pi^2}\bigg\{ \!m_i^2(\phi)\!\left[\log\!\left(\frac{m_i^2(\phi)}{m_i^2(v)}\right) \!-\!\frac32\right]\!+\! 2 m_i^2(v)\! \bigg\}\nn\\
&& +\ \frac{T^4}{2\pi^2}\sum_i n_i \int_0^\infty dx\,x^2 \log\left(1\mp e^{-\sqrt{{m_i^2(\phi)}/{T^2}+x^2}}\right)\nn\\
&& + \ \frac{T}{12\pi} \sum_j n'_j \left\{m_j^3(\phi)- \left[m_j^2(\phi) + \Pi_j(T)\right]^{\frac32}\right\} \ .
\eea
In the expression above the sums are over all particles charged under the ${\rm U}(1)$ including the Goldstone bosons $\chi_{\rm GB}$, $m_i(\phi)$ are the field-dependent masses, $n_i$ is the number of degrees of freedom for a given particle ($n_{Z'}=3$, $n_\phi = 1$, $n_{\chi_{\rm GB}} = 1$), $n_j'$ is similar but includes only scalars and  longitudinal components of vector bosons  ($n'_{Z'}=1$, $n'_\phi = 1$, $n'_{\chi_{\rm GB}} = 1$), and for the Goldstones one needs to replace $m_{\chi_{\rm GB}}(v) \to m_\phi(v)$.
We will assume that  all new Yukawa couplings are small, so that the only relevant field-dependent masses are
\bea\label{31e}
&&m_{Z'}(\phi) = x g \phi \ , \ \ \ \ m_\phi(\phi) = \sqrt{\lambda (3\phi^2-v^2)} \ , \nn\\
&&m_{\chi_{\rm GB}}(\phi) = \sqrt{\lambda (\phi^2-v^2)} \ .
\eea
In the limit $\lambda \ll g$, the thermal masses are
\bea\label{thermalm}
\Pi_{Z'}(T) &=& \frac{1}{3} \left(x^2+\frac92\right)g^2 T^2 \ , \nn\\
 \Pi_\phi(T) &=& \Pi_{\chi_{\rm GB}} (T) = \frac14 x^2 g^2 T^2 \ ,
\eea
where for gauged baryon number $g=g_B$ and $x=3$, while for gauged lepton number $g=g_L$ and $x=2$.

\begin{figure}[t!]
\includegraphics[trim={1.5cm 0.8cm 1cm 0cm},clip,width=8.8cm]{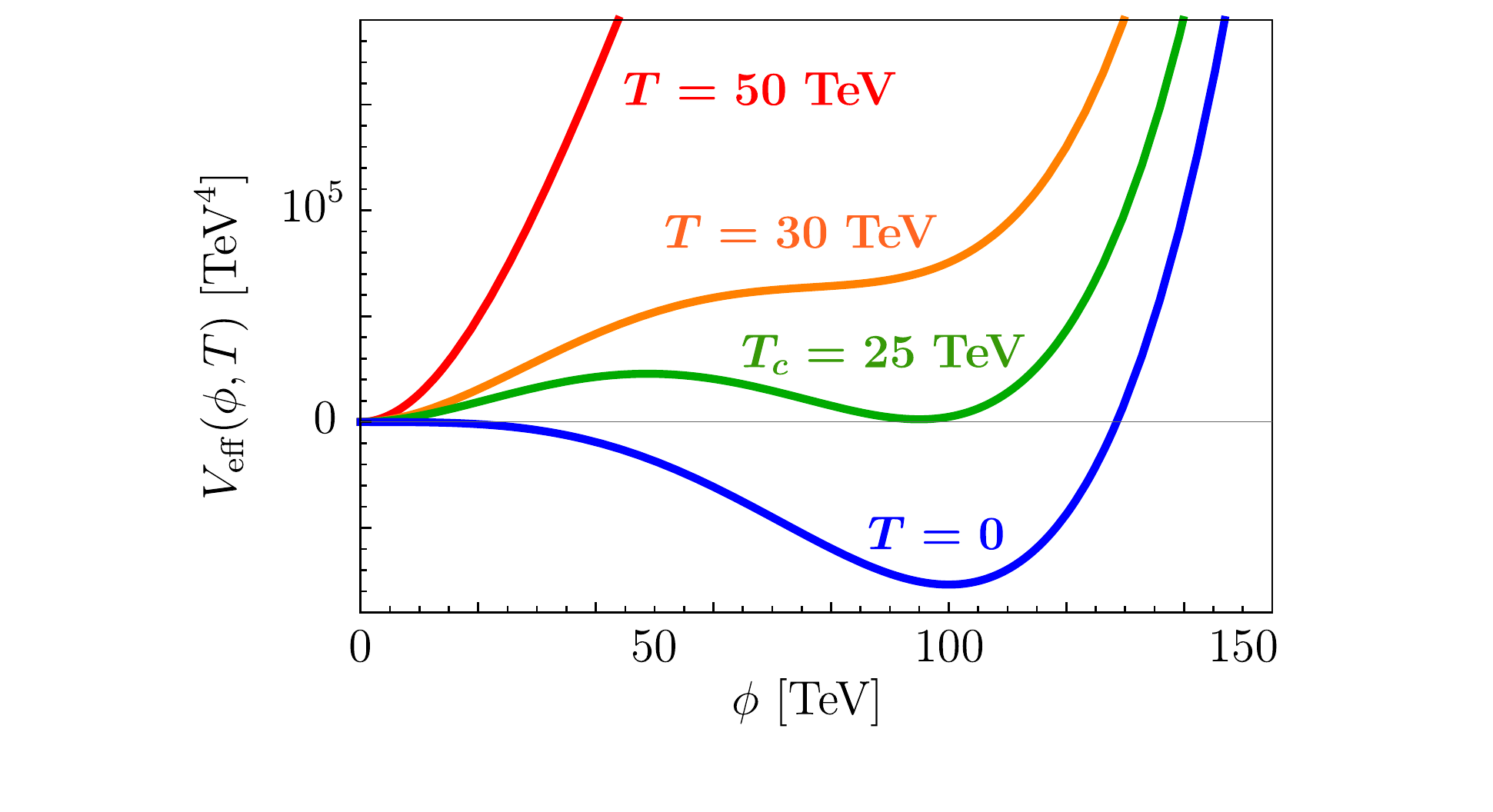} \vspace{-6mm}
\caption{The effective potential of the  model,  $V_{\rm eff}(\phi_B, T)$, plotted for $v_B = 100 \ \rm TeV$, $g_B = 0.25$, $\lambda_B = 0.006$, and several temperatures.}\label{fig:5}
\end{figure}

For a range of $\lambda$ and $g$ values the effective potential develops a vacuum at $\phi \ne 0$ (true vacuum) with a lower energy density than the high temperature vacuum at $\phi=0$ (false vacuum), separated by a potential bump, which are precisely the conditions needed for a first order phase transition to take place. The changing shape of the effective potential is shown in Fig.\,\ref{fig:5} for a particular choice of parameters, in the case of gauged baryon number.

\subsection*{Dynamics of the phase transition}

A first order phase transition from the false vacuum to the true vacuum of a given patch of the Universe corresponds to the nucleation of a  bubble which then starts expanding. This process is initiated at the 
 nucleation temperature $T_*$, which is determined from the condition that the bubble nucleation rate  \cite{LINDE1983421} becomes comparable with the Hubble expansion,
 \bea\label{nucl}
 \bigg(\frac{S(T_*)}{2\pi T_*}\bigg)^{3/2}T_*^4 \,e^{-{S(T_*)}/{T_*}}  \approx H(T_*)^4 \ ,
\eea
where  $S(T)$ is the Euclidean action given by
\bea
S(T)= \int d^3 r \left[\frac12 \left(\frac{d \phi_b}{dr}\right)^2+V_{\rm eff}(\phi_b, T)\right] , 
\eea
with $\phi_b$ being the solution of the bubble equation,
\bea
\frac{d^2 \phi}{dr^2}+\frac{2}{r}\frac{d\phi}{dr}-\frac{d V_{\rm eff}(\phi,T)}{d\phi} = 0 \ ,
\eea
subject to the boundary conditions 
\bea
\frac{d\phi}{dr}\bigg|_{r=0} = 0 \ , \ \ \ \ \ \phi(\infty) = \phi_{\rm false} \ .
\eea
Since $H(T) \approx (T^2 / M_{Pl})\sqrt{4\pi^3 g_*/45}$, Eq.\,(\ref{nucl}) becomes
\bea\label{ST444}
\frac{S(T_*)}{T_*}   \approx  4\log\!\left(\frac{M_{Pl}}{T_*}\right)  \!-\! \log\left[\left(\frac{4\pi^3g_*}{45}\right)^{\!\!2}\!\left(\frac{2\pi \,T_*}{S(T_*)}\right)^{\!\!\frac32}\right]\!. \  \ 
\eea

The phase transition parameters relevant for determining  the gravitational wave signal are: the bubble wall velocity $v_w$, the nucleation temperature $T_*$, the phase transition  strength $\alpha$, and its duration $1/\tilde\beta$. In our analysis we assume $v_w = c$, but other choices are also possible \cite{Espinosa:2010hh,Caprini:2015zlo}. The other three parameters, $T_*$, $\alpha$, and $\tilde\beta$, are determined from the behavior of the effective potential with changing temperature. As such, those parameters encode information about the details of the particle physics model considered.

The phase transition strength is calculated as the ratio of the energy density difference between the false and true vacuum, and that of radiation, both taken at nucleation temperature,
\bea\label{alf}
\alpha = \frac{\rho_{\rm vac}(T_*)}{\rho_{\rm rad}(T_*)} \ .
\eea
Those two quantities are obtained from the relations
\bea
\rho_{\rm vac}(T) &=& V_{\rm eff}(\phi_{\rm false},T) -  V_{\rm eff}(\phi_{\rm true},T)\nn\\
&-&T \frac{\partial}{\partial T} {\left[ V_{\rm eff}(\phi_{\rm false},T) -  V_{\rm eff}(\phi_{\rm true},T)\right]} \ , \nn\\
\rho_{\rm rad}(T) &=& \frac{\pi^2}{30} g_* T^4 \ ,
\eea
where $g_*$ is the number of degrees of freedom active when the bubbles are nucleated. The parameter $\tilde\beta$, related to the time scale of the phase transition, is determined via
\bea\label{bet}
\tilde{\beta} = T_* \frac{d}{dT} \bigg(\frac{S(T)}{T}\bigg)\bigg|_{T=T_*} \ .
\eea

\subsection*{Gravitational wave spectrum}

The dynamics of the nucleated bubbles generates gravitational waves through sound shock waves in the early Universe plasma, bubble collisions, and magnetohydrodynamic turbulence. The expected contribution of each of those processes to the stochastic gravitational wave background was determined through numerical simulations, and the corresponding empirical formulas were derived. The resulting gravitational wave signal is the sum of the three contributions,
\bea
h^2 \Omega_{\rm PT}(f) = h^2\Omega_s(f) + h^2\Omega_c(f) + h^2\Omega_t(f) \ . 
\eea
The expected signal from sound waves  is  \cite{Hindmarsh:2013xza,Caprini:2015zlo}
\bea\label{sound}
h^2 \Omega_{s}(f) \,&\approx&\, \frac{1.9\times 10^{-5}}{\tilde\beta}\left(\frac{\kappa_s\,\alpha}{\alpha+1}\right)^2\left(\frac{100}{g_*}\right)^{\frac13} \Upsilon\nn\\
&\times&\frac{(f/f_s)^3}{\big[1+0.75 (f/f_s)^2\big]^{7/2}}  \ ,
\eea
where $f_s$ is the peak frequency, $\kappa_s$ is the  fraction of the latent heat transformed into the bulk motion of the plasma \cite{Espinosa:2010hh}, and $\Upsilon$ is  the suppression factor \cite{Ellis:2020awk,Guo:2020grp},
\bea\label{sp}
f_s &=& (0.19 \ {\rm Hz} ) \left(\frac{T_*}{1 \ {\rm PeV}}\right)\left(\frac{g_*}{100}\right)^\frac16  \tilde\beta \ ,\nn\\
\kappa_s &=& \frac{\alpha}{0.73+0.083\sqrt\alpha + \alpha} \ ,\nn\\
\Upsilon\, &=& 1- \frac1{\Big[1+\frac{8\pi^{\frac13}}{\tilde{\beta}}\big(\frac{{\alpha+1}}{3\kappa_s\alpha}\big)^{\frac12}\Big]^{\frac12}} \ .
\eea
The signal from bubble wall collisions is \cite{Kosowsky:1991ua,Huber:2008hg,Caprini:2015zlo} (see \cite{Lewicki:2020azd} for recent updates)
\bea\label{colll}
h^2 \Omega_{c}(f) \,&\approx&\, \frac{4.9\times 10^{-6}}{\tilde\beta^2}\left(\frac{\kappa_c \, \alpha}{\alpha+1}\right)^2\left(\frac{100}{g_*}\right)^{\frac13}\nn\\
&\times&\frac{(f/f_c)^{2.8}}{1+2.8 (f/f_c)^{3.8}} \ , \ \ \ \ \ \ 
\eea
where now $f_c$ is the peak frequency and $\kappa_c$ is the fraction of the latent heat deposited into the bubble front \cite{Kamionkowski:1993fg},
\bea
f_c &=& (0.037 \ {\rm Hz} ) \left(\frac{T_*}{1 \ {\rm PeV}}\right)\left(\frac{g_{*}}{100}\right)^\frac16\tilde\beta \ , \nn\\
\kappa_c &=& \frac{\frac{4}{27}\sqrt{\frac32\alpha}+0.72\,\alpha}{1+0.72 \,\alpha} \ .
\eea
The final contribution is provided by turbulence  \cite{Caprini:2006jb,Caprini:2009yp},
\bea\label{tur}
h^2 \Omega_{t}(f) \,&\approx&\, \frac{3.4\times 10^{-4}}{\tilde\beta}\left(\frac{\epsilon\,\kappa_s \, \alpha}{\alpha+1}\right)^{\frac32}\left(\frac{100}{g_*}\right)^{\frac13} \nn\\
&\times&\frac{ ({f}/{f_t})^{3}}{\big(1+{8\pi f}/{f_*}\big)\big(1+{f}/{f_t}\big)^{{11}/{3}}}\ ,
\eea
where $\epsilon=0.05$ \cite{Caprini:2015zlo}, while the  peak frequency $f_t$ and the parameter $f_*$ are
\bea
f_t &=& (0.27 \ {\rm Hz} ) \left(\frac{g_*}{100}\right)^\frac16\left(\frac{T_*}{1 \ {\rm PeV}}\right) \tilde\beta\ ,\nn\\
f_* &=& (0.17 \ {\rm Hz})\left(\frac{g_*}{100}\right)^\frac16\left(\frac{T_*}{1 \ {\rm PeV}}\right) \ .
\eea

\begin{figure}[t!]
\includegraphics[trim={2.4cm 0.8cm 1cm 0cm},clip,width=9.5cm]{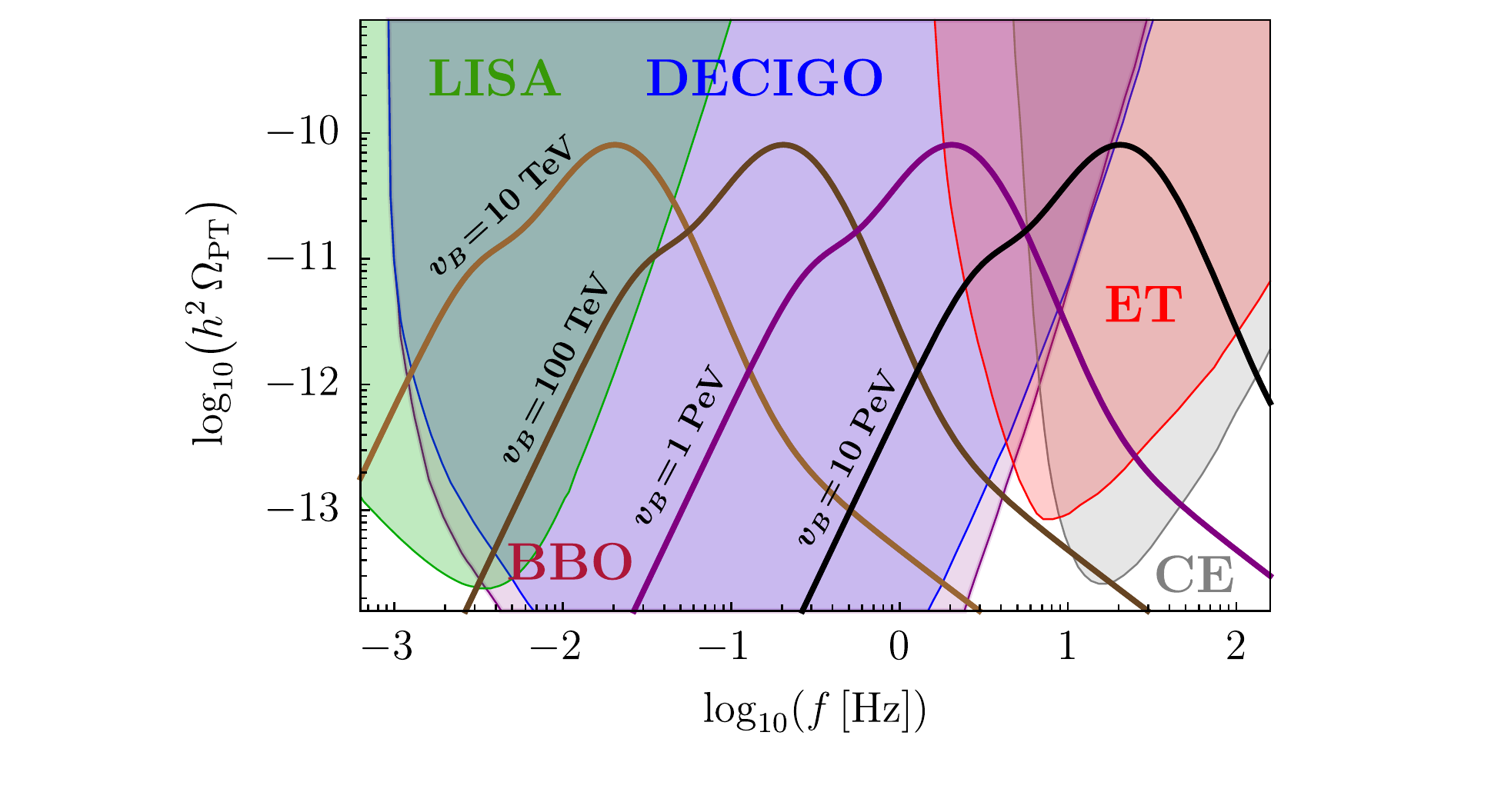} \vspace{-7mm}
\caption{Stochastic gravitational wave background from first order phase transitions triggered by ${\rm U}(1)_B$ breaking assuming the model parameters $g_B = 0.25$ and $\lambda_B = 0.006$, for several choices of the symmetry breaking scale. The shaded regions correspond to the sensitivity of future gravitational wave detectors, as in Fig.\,\ref{fig:1}.}\label{fig:6}
\end{figure}

The resulting gravitational wave signals from the first order phase transition triggered by ${\rm U}(1)_B$ breaking  are shown in Fig.\,\ref{fig:6} for several symmetry breaking scales: $10 \ {\rm TeV}$ (light brown curve), $100 \ {\rm TeV}$ (brown curve), $1 \ {\rm PeV}$ (purple curve), and $10 \ {\rm PeV}$ (black curve). The gauge  coupling was assumed to be $g_B = 0.25$, and the quartic coupling was chosen to be $\lambda_B=0.006$. Spectra with peaks at larger frequencies  correspond to higher symmetry breaking scales.

The effect of the suppression factor $\Upsilon$ reducing the sound wave contribution in Eq.\,(\ref{sound}) is that the bubble collision and turbulence components become distinguishable in the spectrum. Although the main peak is still due to sound waves, the slope at lower frequencies is dominated by bubble collisions, whereas for higher frequencies the turbulence contribution visibly changes the slope of the curve. Without the suppression factor, the spectrum would be determined, to a good approximation, just by the sound wave component in the region relevant for future  detectors.  

\begin{figure}[t!]
\includegraphics[trim={2.4cm 0.8cm 1cm 0cm},clip,width=9.0cm]{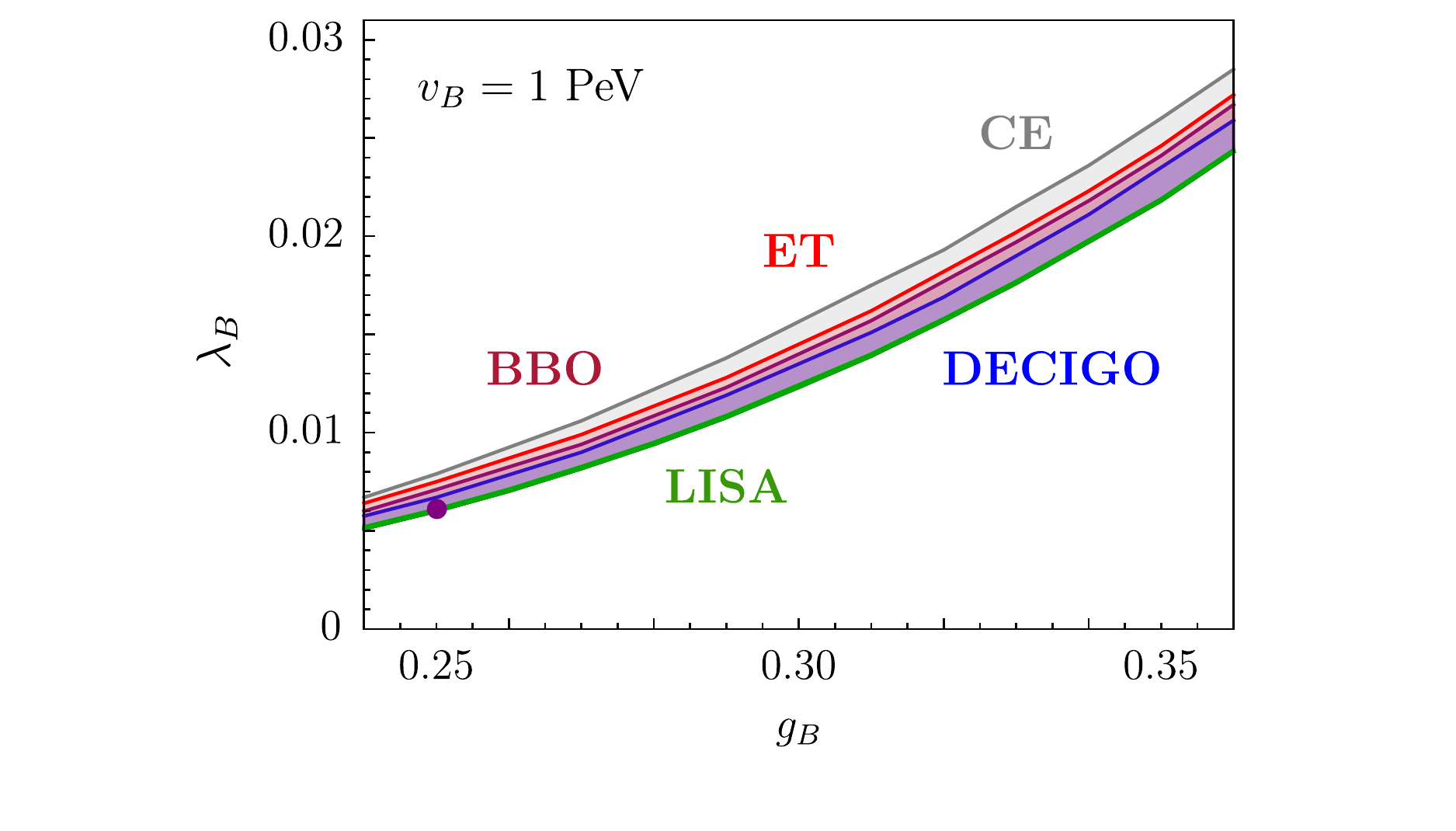} \vspace{-7mm}
\caption{Regions of parameter space $(g_B, \lambda_B)$ assuming $v_B = 1 \ {\rm PeV}$ where the gravitational wave signal from a first order phase transition has a signal-to-noise ratio  greater than five upon one year of data taking in various experiments. The choice of colors matches that in Fig.\,\ref{fig:6}, and the dot corresponds to the purple curve.}\label{fig:7}
\end{figure}

Depending on the  Lagrangian parameters, the signal may be detectable in upcoming gravitational wave experiments: LISA, DECIGO, Big Bang Observer, Einstein Telescope, and Cosmic Explorer. To study this more quantitatively, in Fig.\,\ref{fig:7} we show part of the   $(g_B, \lambda_B)$ parameter space for which the signal can be detected in each experiment when $v_B=1 \ {\rm PeV}$. Specifically, the upper boundary for each detector corresponds to  a signal-to-noise ratio of five upon a single year of data taking, while the lower boundary  arises  when either $ S(T)/T$ is too large to satisfy Eq.\,(\ref{ST444}) or the new vacuum has an energy density larger than that of the high temperature vacuum.

As mentioned earlier, our analysis of the first order phase transition signal from ${\rm U}(1)_B$ breaking  differs from the one in \cite{Fornal:2020esl} in several aspects. The fermionic particle content  in Eq.\,(\ref{fermions}) is different, and we chose the corresponding Yukawa couplings to be small, which is a more minimal scenario than  $Y=0.6$ in  \cite{Fornal:2020esl}. Our analysis is also applicable to  ${\rm U}(1)_L$ breaking, since we calculate the thermal masses in the general case -- this result will be used in Section \ref{signatures}. Finally, when determining  the gravitational wave signal we included the effect of bubble collisions, which was not considered in \cite{Fornal:2020esl}, but which increases the reach of upcoming detectors due to the enhancement of the signal in the lower frequency region.

\section{Gravitational wave signatures}
\label{signatures}

In this section we demonstrate the  diversity of gravitational wave signatures expected within the  framework of the model, searchable in near-future experiments. The  cases enumerated in Section \ref{modell}, corresponding to the possible structures of the scalar sectors, give rise to the coexistence in the spectrum of the following  gravitational wave signals from first order phase transitions (PT), cosmic strings (CS), and domain walls (DW):
\begin{itemize}
\item[$(a)$] ${\rm (PT + PT),  (PT + CS), (CS + CS);}$
\item[$(b,c)$] ${\rm (PT + PT), (PT + CS), (PT + DW), (CS + DW);}$
\item[$(d)$] ${\rm (PT + PT), (PT + DW), (DW + DW).}$
\end{itemize}
We discuss below explicit examples of how  those signatures are realized in our model. Two of them, ${\rm (DW + DW)}$ and  ${\rm (CS + DW)}$ have not been considered in the literature before, whereas  ${\rm (PT + CS)}$, ${\rm (PT + PT)}$  and  ${\rm (PT + DW)}$ have  been already proposed. The case ${\rm (CS + CS)}$ does not give rise to any new features, since the  signal is  dominated by the cosmic string contribution from the higher symmetry breaking due to the flatness of the cosmic string spectrum.

\subsection*{Domain walls + domain walls}

\begin{figure}[t!]
\includegraphics[trim={2.4cm 0.8cm 1cm 0cm},clip,width=9.5cm]{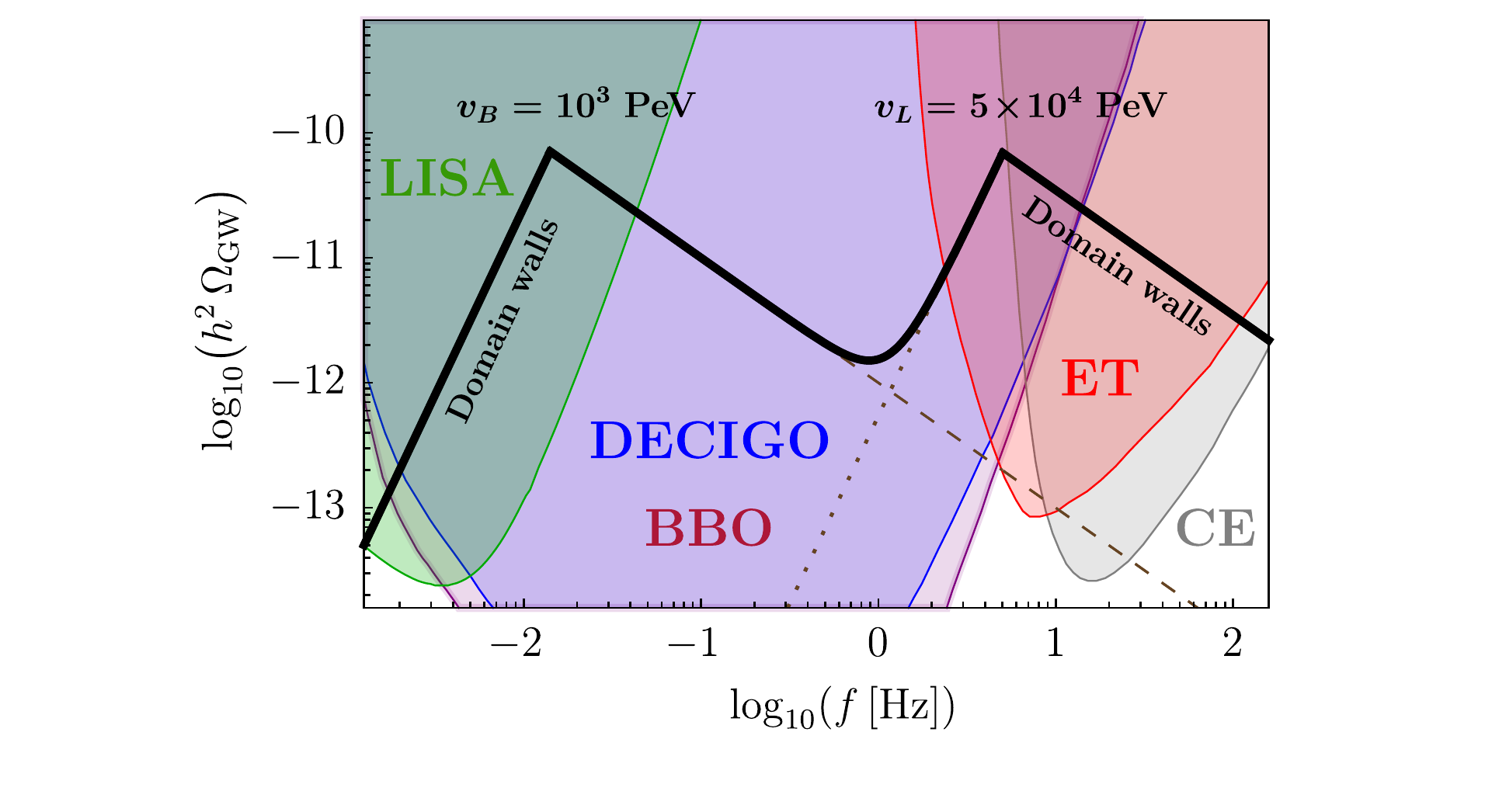} \vspace{-8mm}
\caption{First novel gravitational wave signature of the model consisting of a double domain wall peak, realized when each of the  ${\rm U}(1)$  symmetries is  broken by two scalars --  case $(d)$. \\ }\label{fig:8}
\end{figure}

\begin{figure}[t!]
\includegraphics[trim={2.4cm 0.8cm 1cm 0cm},clip,width=9.5cm]{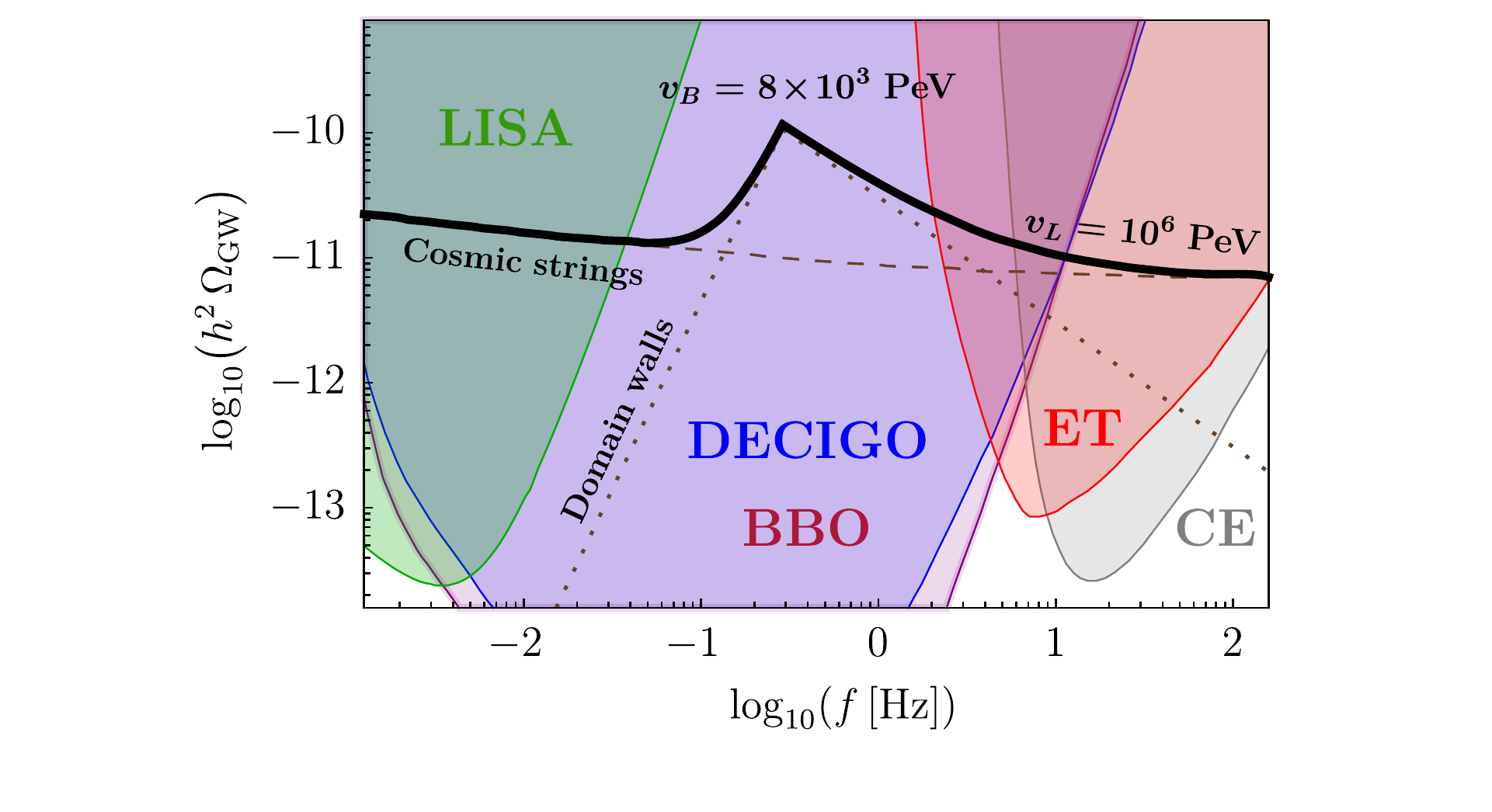} \vspace{-8mm}
\caption{Second novel gravitational wave signature consisting of a domain wall peak over a cosmic string background, realized when one ${\rm U}(1)$ symmetry is broken by one scalar and the other ${\rm U(1)}$ is broken by two scalars -- cases $(b)$ and $(c)$. }\label{fig:9}
\end{figure}

A new gravitational wave signature arises when each of the two ${\rm U}(1)$ symmetries is broken by two scalars, leading to the production of domain walls at two different energy scales during the evolution of the Universe. The signal consists of two sharp domain wall peaks. The slope on the left side of each peak depends on the frequency like $\sim f^3$, whereas the slope on the right side of the peak falls like $\sim 1/f$. There is a nontrivial structure created between the two peaks, which can be used to distinguish this type of signal from others. If the two symmetry breaking scales are high, this signature can be searched for in all the upcoming gravitational wave experiments we considered: LISA, DECIGO, Big Bang Observer, Einstein Telescope, and Cosmic Explorer. 
A realization of this scenario in our model is shown in Fig.\,\ref{fig:8}, where the parameters for the ${\rm U(1)}_B$ breaking were chosen to be $v_B =10^3\ {\rm PeV}$ and $\Delta \rho = 10^{-5}\ {\rm PeV^4}$, whereas for the ${\rm U(1)}_L$ breaking they are $v_L =  5\times10^4 \ {\rm PeV}$ and $\Delta \rho = 1.6\times 10^5\ {\rm PeV^4}$.

\subsection*{Cosmic strings + domain walls}

Another gravitational wave signature, not considered in the literature before, is realized when one of the ${\rm U}(1)$ symmetries is broken by one scalar, leading to cosmic string production, whereas the other ${\rm U}(1)$ is broken by two scalars, resulting in domain wall creation. If the two symmetry breaking scales are  high, their contributions may overlap and produce a very unusual domain wall peak over the cosmic string background. An example is shown in Fig.\,\ref{fig:9}, where the symmetry breaking scale for ${\rm U}(1)_L$ was chosen to be  $v_L = 10^{6} \ {\rm PeV}$, whereas the parameters for ${\rm U}(1)_B$ breaking are $v_B= 8\times10^3 \ {\rm PeV}$ and $\Delta \rho = 2 \ {\rm PeV^4}$. For this particular selection of parameters, Big Bang Observer and DECIGO can probe the peak area, but for lower ${\rm U}(1)_B$ breaking scales this structure is accessible to LISA, whereas higher symmetry breaking scales would make it detectable by  Einstein Telescope and Cosmic Explorer.

\begin{figure}[t!]
\includegraphics[trim={2.4cm 0.8cm 1cm 0cm},clip,width=9.5cm]{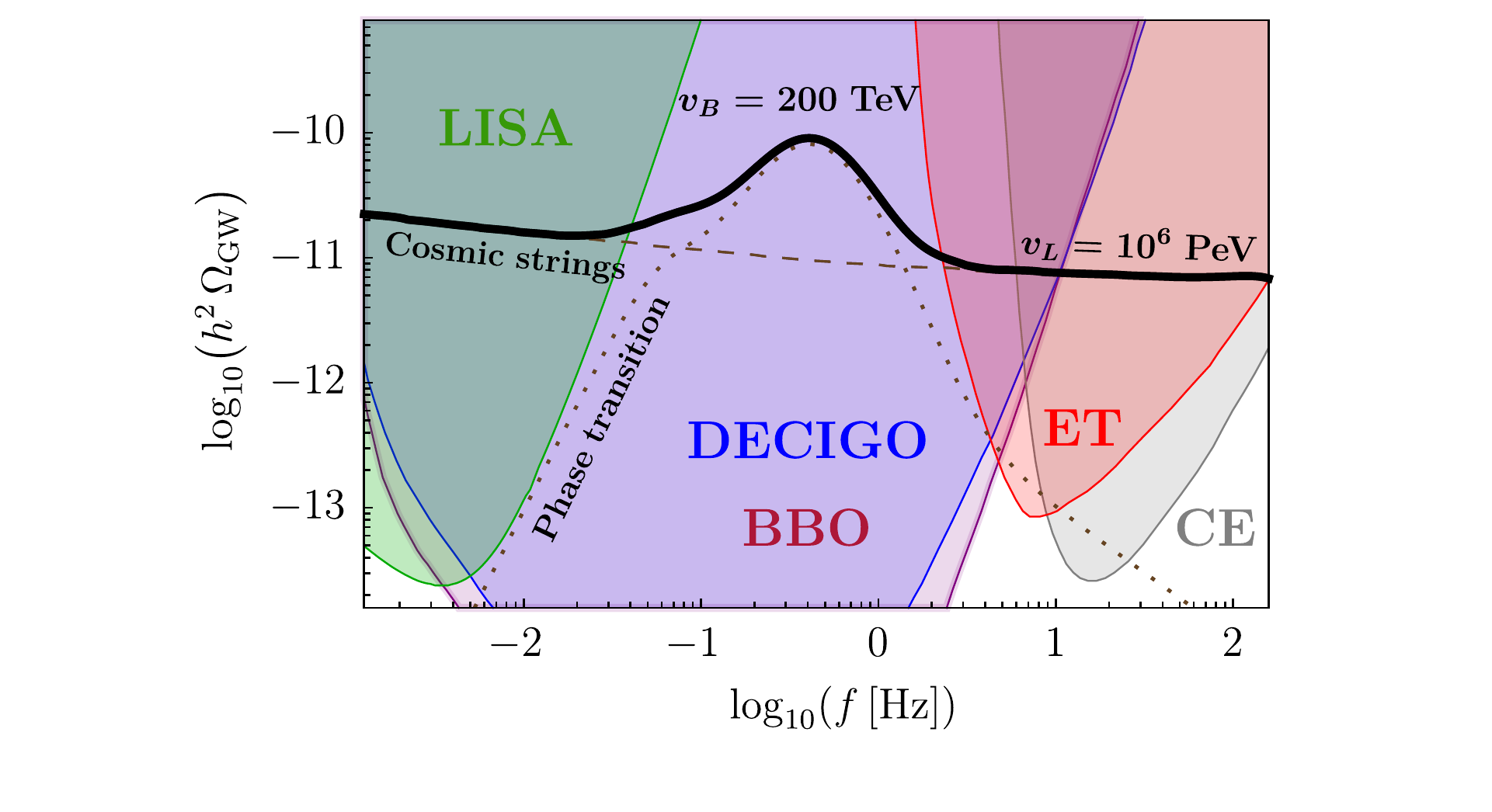} \vspace{-8mm}
\caption{Gravitational wave signature with a first order phase transition peak over a cosmic string background, first proposed in \cite{Fornal:2020esl}, realized when one  ${\rm U}(1)$ is broken by only one scalar and the other ${\rm U}(1)$ is broken either by one or two scalars -- cases $(a)$, $(b)$, $(c)$.}\label{fig:10}
\end{figure}

\subsection*{Phase transition + cosmic strings}

If one of the symmetries is broken by a single scalar at the high scale, and the other symmetry is broken by either one or two scalars at the low scale, this can lead to a gravitational wave signature consisting of a phase transition bump over a cosmic string background. This signature was first proposed in \cite{Fornal:2020esl} in the context of a different gauged baryon and lepton number model with a well-motivated large hierarchy between symmetry breaking scales, and recently considered in another scenario \cite{Ferrer:2023uwz}. In Fig.\,\ref{fig:10} we show an example of such a signal, where ${\rm U}(1)_L$ is broken at the scale $v_L=10^6 \ {\rm PeV}$, whereas the scale of ${\rm U}(1)_B$ breaking is $v_B=200 \ \rm TeV$. The other parameter values for this particular plot are $g_B=0.25$ and $\lambda_B=0.006$. As in the previous case, by changing the scale of ${\rm U}(1)_B$ breaking the  bump can shift and become searchable not only by Big Bang Observer and DECIGO, but also by Cosmic Explorer, Einstein Telescope, or LISA.

\subsection*{Phase transition + phase transition}

\begin{figure}[t!]
\includegraphics[trim={2.4cm 0.8cm 1cm 0cm},clip,width=9.5cm]{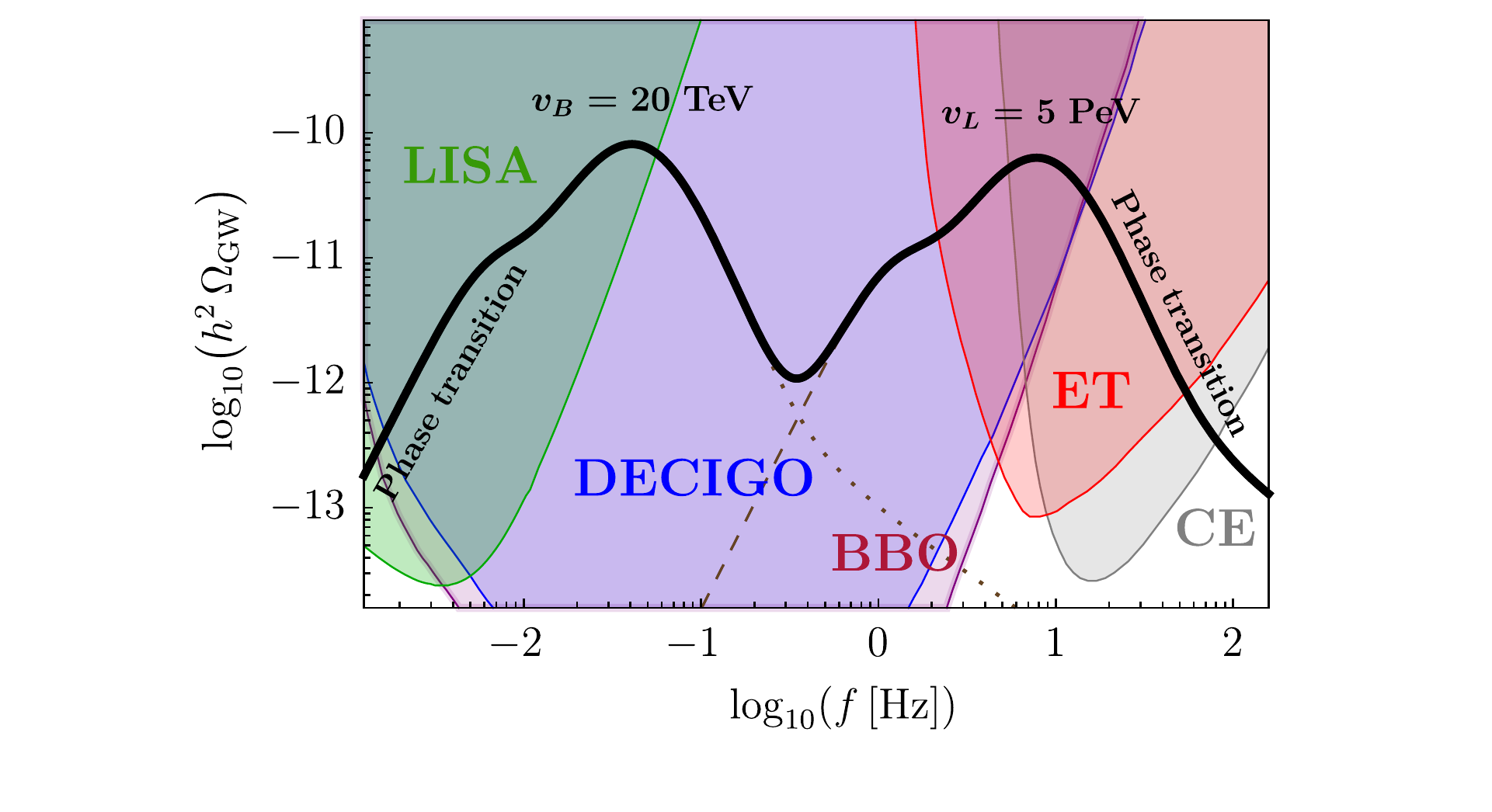} \vspace{-6mm}
\caption{Gravitational wave signature consisting  of two first order phase transition peaks, similar to the ones proposed in \cite{Angelescu:2018dkk,Greljo:2019xan,Fornal:2020ngq}, arising when the two ${\rm U}(1)$ symmetries are broken by any number of scalars -- realized in cases $(a)$, $(b)$, $(c)$, $(d)$.}\label{fig:11}
\end{figure}

Independently of the scalar sector structure, the breaking of  two ${\rm U}(1)$ gauge symmetries  can always result in a gravitational wave signal with two first order phase transition peaks. Such a signature is generically expected in theories with a multistep symmetry breaking pattern, and has been proposed for various  models of new physics \cite{Angelescu:2018dkk,Greljo:2019xan,Fornal:2020ngq}. In Fig.\,\ref{fig:11} a realization of this signature is shown in the case of our model, assuming that the ${\rm U}(1)_B$ symmetry is broken by one scalar at  the scale $v_B = 20 \ {\rm TeV}$ (the other parameters are $g_B=0.25$ and $\lambda_B=0.006$), and  the  ${\rm U}(1)_L$ symmetry is broken also by one scalar at  the scale $v_L = 5 \ {\rm PeV}$ (with  $g_L=0.20$ and $\lambda_L=0.0025$). We note that for the two contributions  appropriate formulas for the thermal masses were adopted, according to Eq.\,(\ref{thermalm}). Such a signal can be searched for in all the future gravitational wave detectors we considered.

\subsection*{Phase transition + domain wall}

\begin{figure}[t!]
\includegraphics[trim={2.4cm 0.8cm 1cm 0cm},clip,width=9.5cm]{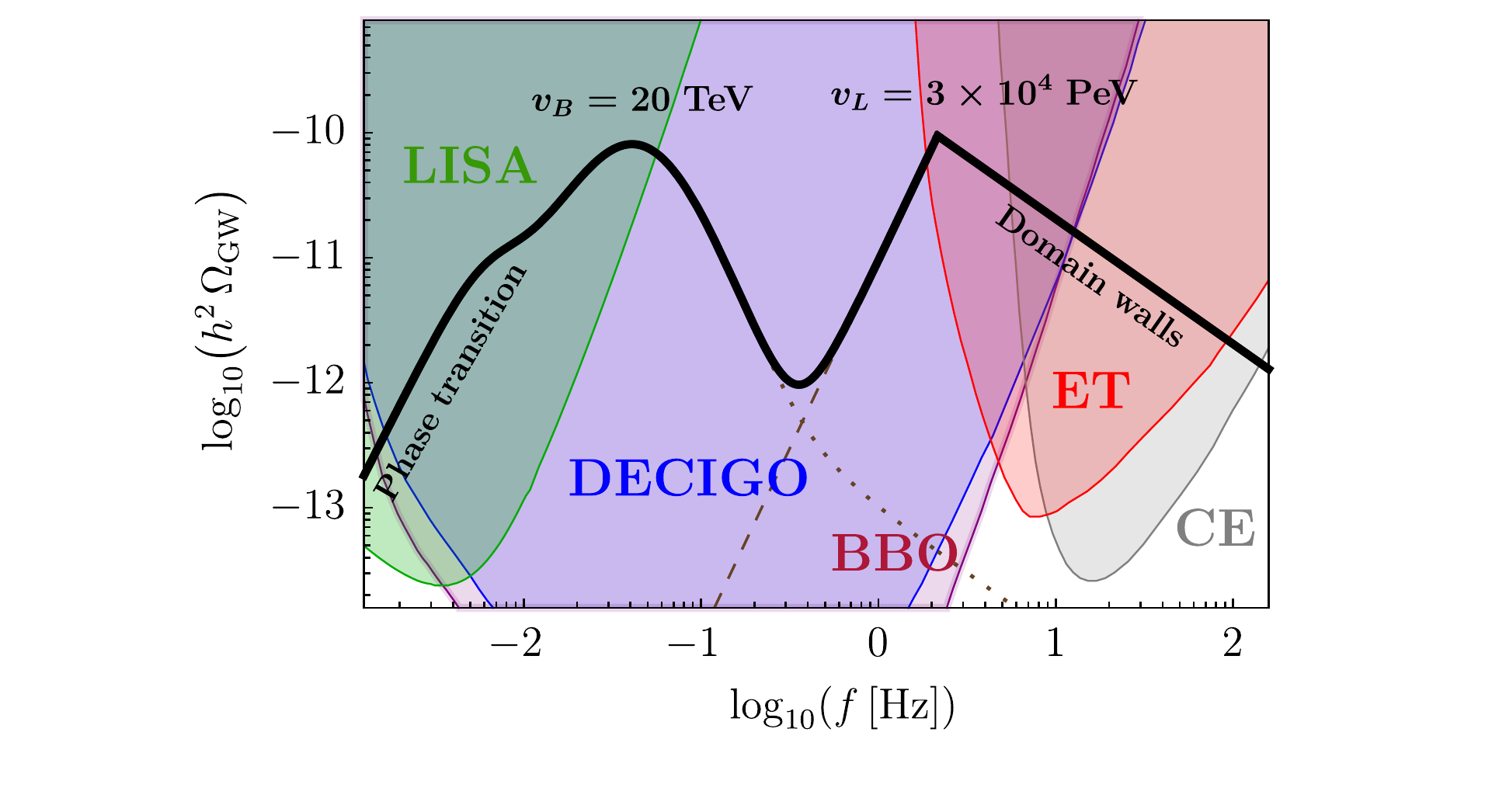} \vspace{-6mm}
\caption{Gravitational wave signature containing a phase transition peak and a domain wall peak, proposed in \cite{Fornal:2023hri}, realized when at least one ${\rm U}(1)$ symmetry  is broken by two scalars -- cases $(b)$, $(c)$, $(d)$. }\label{fig:12}
\end{figure}

The final qualitatively different signature consists also of two peaks, but this time one coming  from a first order phase transition and the second one arising from domain wall annihilation. Such a signal has very recently been proposed in  \cite{Fornal:2023hri}. In our model it can be realized if there is a large hierarchy between the ${\rm U}(1)_B$ and ${\rm U}(1)_L$ symmetry breaking scales. Figure {\ref{fig:12}} shows a realization of this scenario when 
 ${\rm U}(1)_L$ is broken by two scalars at the scale $v_L = 3\times10^4 \ {\rm PeV}$ (with a potential bias $\Delta\rho = 6.3\times10^3 \ {\rm PeV^4}$), whereas ${\rm U}(1)_B$ is broken by one scalar at the scale $v_B = 20 \ {\rm TeV}$ (with the other parameters being $g_B=0.25$ and $\lambda_B=0.006$). As pointed out in \cite{Fornal:2023hri}, the two peaks may appear in a different order, which would happen for a ${\rm U}(1)_L$ breaking scale of $v_L \sim 10^3 \ {\rm PeV}$ and a ${\rm U}(1)_B$ breaking scale of $v_B \sim 10 \ {\rm PeV}$. In both scenarios, the signature can be searched for in the upcoming gravitational wave detectors we focused on.

Although the signatures discussed above can be realized for any pattern of symmetry breaking, the phenomenologically more attractive scenarios involve ${\rm U}(1)_L$ broken at the high scale, so that the bound in Eq.\,(\ref{lepbound}) is satisfied and the theory can successfully accommodate leptogenesis, as discussed in Section \ref{dam}. Additionally, with no motivation for a low ${\rm U}(1)_B$ breaking scale in the model we are considering, the new signatures shown in Figs.\,\ref{fig:8} and \ref{fig:9} can be naturally realized, and are quite appealing given the reach of the upcoming gravitational wave experiments.

\section{Conclusions and Outlook}

It is truly extraordinary that gravitational wave astronomy can join forces with elementary particle physics to search for answers to fundamental questions about the structure of the Universe and its earliest stages of evolution. Indeed, processes happening at energies too high to be probed by conventional particle physics detectors (such as high scale leptogenesis and seesaw mechanism) can leave a remarkable imprint through the primordial  gravitational wave background emitted soon after the Big Bang. Detecting such a signal would bring us closer to discovering which, if any,  of the proposed Standard Model extensions addressing the outstanding questions about dark matter, matter-antimatter asymmetry, or neutrino masses, is realized in Nature.

A stochastic gravitational wave background is expected to originate in the early Universe within the framework of many particle physics models through first order phase transitions, cosmic string dynamics  and domain wall annihilation. In particular, explaining the matter-antimatter asymmetry puzzle requires a first order phase transition to happen, indicating the huge importance of  stochastic gravitational wave searches. The literature referred to in Section \ref{intr} contains analyzes of such signatures in theories beyond the Standard Model, however, the majority of the works focus on one single component at a time, generally not looking at the possible interplay between the contributions  from different sources.

In this paper we highlighted the importance of searches for novel gravitational wave signatures arising when multiple components are present in the spectrum and add up producing new features in the signal. Such unique signatures are expected  in theories with  more than one symmetry breaking, and result from the interplay between the contributions from first order phase transitions, cosmic strings, and/or domain walls. The new gravitational wave signals we propose to look for are: $(1)$ Double-sharp-peak structure from domain walls produced when  two gauge symmetries are broken by multiple scalars; $(2)$ Domain wall peak over a cosmic string plateau when one symmetry is broken by a single scalar and the other symmetry is broken by multiple scalars.

Although we demonstrate  how those signatures arise in a specific model with gauged baryon and lepton number, our results are applicable to a much wider class of theories with two ${\rm U}(1)$ gauge symmetries broken at different energy scales. Indeed, the new signals consist of the
 cosmic string and domain wall contributions, thus they are fairly model-independent, since the cosmic string component depends only on the symmetry breaking scale, whereas the domain wall contribution depends  on the symmetry breaking scale and the potential bias. Our results can also be extended to models with non-Abelian gauge groups. As already suggested in \cite{Fornal:2023hri}, it would be interesting to investigate the case when one of the symmetries is ${\rm SU}(2)$ broken by two scalar triplets, as this can result in the production of cosmic strings \cite{Hindmarsh:2016lhy}, and could perhaps lead to new signals involving  contributions from all three processes: first order phase transitions, cosmic string dynamics, and domain wall annihilation.

The gravitational wave signatures discussed here can be searched for in upcoming experiments, including LISA, Big Bang Observer, DECIGO, Cosmic Explorer, and Einstein Telescope, 
enabling  those detectors to probe the structure of high-scale symmetry breaking sectors. This is especially relevant for theories of leptogenesis such as the model we considered, in which, contrary to \cite{Duerr:2013dza,Fornal:2020esl}, the scale of ${\rm U(1)_B}$ symmetry breaking is not bounded from above and can also be high, allowing for signals $(1)$ and $(2)$ to be generated.

Finally, it is worth mentioning that a spontaneous breaking of a single gauge symmetry can by itself lead to gravitational wave signatures combining signals from a phase transition and cosmic strings, or a  phase transition and domain walls.  Given the sensitivity of the experiments considered, the symmetry breaking scale would have to be $\sim 100 - 1000 \ {\rm PeV}$ for the combined signal to be discoverable. The cosmic string contribution would then be detectable by Big Bang Observer and DECIGO, the phase transition peak could be seen by Cosmic Explorer and Einstein Telescope, and the domain wall peak  would be visible in LISA. Investigating this in more detail is an interesting follow-up project, and could be tied to gravitational wave experiments sensitive to lower frequencies, such as the pulsar timing arrays: NANOGrav \cite{NANOGRAV:2018hou}, PPTA \cite{2013PASA...30...17M}, EPTA \cite{2010CQGra..27h4014F}, IPTA \cite{2010CQGra..27h4013H}, or SKA \cite{Weltman:2018zrl}.

\vspace{5mm}

\subsection*{Acknowledgments}

This research was supported by the National Science Foundation under Grant No. PHY-2213144.

\bibliography{bibliography}

\end{document}